\documentclass[journal]{IEEEtai}

\usepackage[colorlinks,urlcolor=blue,linkcolor=blue,citecolor=blue]{hyperref}
\usepackage{color,array}
\usepackage{graphicx}
\usepackage{siunitx}
\usepackage{bm}
\usepackage{float}
\usepackage{stfloats}
\usepackage{amsmath}
\usepackage{amssymb}
\usepackage{amsfonts}
\usepackage{hyperref}
\usepackage{comment}

\begin{document}

\title{Dual-Locking Learned AI Models: A PIN-Based Sparse QIM Watermarking and Adaptive Index Permutation Approach} 

\author{Iva Vasic, \IEEEmembership{Member, IEEE}, Jesús Muñoz-Cádiz, and Bata Vasic
\thanks{Received 30 June 2025; revised 7 October 2025; accepted 18 November 2025. This article was recommended for publication by Associate Editor Kaitai Liang. (\textit{Corresponding author: Iva Vasic.)}
}
\thanks{Iva Vasic and Jesús Muñoz-Cádiz are with the Faculty of Science and Medicine, University of Fribourg, 1700 Fribourg, Switzerland (e-mail: iva.vasic@unifr.ch; jesus.munozcadiz@unifr.ch).}
\thanks{Bata Vasic is with the Faculty of Electronic Engineering, University of Nis, 18000 Nis, Serbia, on leave from the Information Technology Department, University MB, 11040 Belgrade, Serbia (e-mails: bata.vasic@elfak.ni.ac.rs).}
\thanks{Digital Object Identifier 10.1109/TAI.2025.3636862}}

\markboth{Journal of IEEE Transactions on Artificial Intelligence}
{Vasic \MakeLowercase{\textit{et al.}}: Dual-Locking Learned AI models}

\maketitle

\begin{abstract}
We present a dual-locking method for securing trained neural networks, combining key-driven index permutation with PIN-based watermarking via Sparse Quantization Index Modulation (QIM). To incorporate cryptographic randomness into the locking mechanism, we independently apply a uniform random permutation to every row of the adaptively selected index vectors. The second stage embeds a robust, blind binary watermark into the bias coefficients, modulating its quantized values, and binding the network to a user-defined Personal Identification Number (PIN). Without the correct key, the network retains its architecture, but becomes functionally impaired due to disrupted internal representations. Unlocking is achieved through inverse permutation, fully restoring model accuracy. Simultaneously, the embedded watermark remains imperceptible, concealed within the host bias structure without affecting functionality. Upon extraction, it verifies the key association and confirms model authorship. To enhance both robustness and recoverability, we apply an adaptive key selection strategy that redistributes high-magnitude weights to low-sensitivity positions and vice versa. This maximizes degradation when locked, while ensuring full recovery when unlocked. Experiments on MNIST, CIFAR-10/100, and ImageNet-1K using fully connected networks, ResNet Convolutional Neural Networks (CNNs), and transformer architectures demonstrate that our method sharply reduces accuracy when locked to below 10\% and even 0.5\% for CNNs, while fully restoring it with the correct key. The embedded watermark does not degrade performance and reliably authenticates ownership. Embedding analysis across transformers and CNNs reveals the method’s diagnostic potential, where characteristic shifts in embedding distributions may indicate undertraining or suboptimal design. Overall, this approach offers strong protection and traceability for neural network models.
\end{abstract}

\begin{IEEEImpStatement}
This work presents a practical neural networks security solution that enables owners to lock their models through drastic accuracy degradation and prove ownership without harming performance. It is relevant for commercial AI services, government-regulated applications, and academic tools where model theft or tampering could lead to legal, economic, or ethical issues. The system is reversible and verifiable, and by combining two independent security mechanisms, it establishes a new standard for secure AI licensing and distribution. This technology fills a critical gap in model protection and could shape future legal and technological frameworks in AI intellectual property management.
\end{IEEEImpStatement}


\section{Introduction}
\label{sec:I-Introduction}

\IEEEPARstart{T}{oday}, Artificial Intelligence (AI) models are widely employed in almost every industrial area. Companies invest a lot of time and money to improve the performance of their artificial models, designing powerful deep learning and transformers architectures, and providing large training data sets. The result of such investments is clearly shown in enabling many breakthroughs in challenging problems.  The speed and accuracy of problem solving are evident even in public web- and mobile-based applications. An unhampered distribution of such developed models over the whole world market is a new paradigm of modern business. It seems, however, that the accelerated development of AI architecture has suppressed the intellectual property of huge investments and allows a lot of piracy freedom. On the contrary, decades of mass usage of multimedia content have caused a pretty quick development of watermarks and other tools for protecting the intellectual properties of such data distribution. In this sense, the AI training might be considered as a secure process that uses only protected or even encrypted data to achieve a satisfactory network performance. Unfortunately, a lot of training time and knowledge used in architecture design are still vulnerable without adequate protection.

Following an analogy with the best protecting solutions, which are practically proven in the multimedia areas, in the process of AI protection, we can expect a key-dependence of the main protecting features, such as blindness, robustness, and capacity in relation to usability/accuracy of the protected AI model. However, in industry where high accuracy of the AI models is the imperative of large investments, any precision reduction leads to doubt about the protection acceptability. Therefore, the unique representations of the AI structure, whose only primitive elements (weights and biases) can be employed as possible protection carriers, faces us with a new challenge: finding a sustainable protection method without any loss of accuracy.

Fortunately, the Internet-based use of artificial models is typically confined to their deployment in problem-solving tasks, while the commercial distribution of trained models represents a separate, controlled channel. Therefore, unlike multimedia content protection \cite{Vasic2013QIM-LDPC}, which often demands universal and publicly verifiable watermarking \cite{Vasic2019}, neural network protection can leverage scenario-specific techniques. Consequently, a single, all-purpose watermarking approach is neither required nor optimal. From the outset, we deliberately avoided relying exclusively on watermarking techniques, recognizing that private cryptographic methods can provide greater efficiency and robustness when public traceability is not essential.
This paper introduces a dual-locking framework for the protection of learned AI architectures. The first component applies a key-driven encryption scheme that permutes the multidimensional indices of selected weight coefficients, without altering their trained numerical values. Based on the desired key length and encryption strength, the algorithm selects critical weight positions and encodes them into two-dimensional vectors representing the layer index and the intra-layer position of each weight. These vectors are then permuted deterministically using a user-defined key, rendering the network functionally inoperable without proper decryption. The second component introduces a robust PIN-based watermarking stage, embedding a blind binary signature into the bias coefficients via Sparse Quantization Index Modulation (QIM). Although the user only requires the key to unlock the model, this watermark binds the model to a specific Personal Identification Number (PIN), enabling user-specific ownership verification while remaining imperceptible and nonintrusive to the model’s inference performance. Together, these two stages enable both reversible functionality locking and verifiable authorship confirmation, addressing the needs of both network distributors and users.

The rest of the paper is organized as follows. Section \ref{sec:II-Problem Definition and Relevant Work} outlines the AI protection problem and reviews related work, highlighting key contributions and existing limitations. In Section \ref{sec:III-Key-based Permutation Locking Mechanism} we introduce the key-based permutation locking mechanism in detail, presenting its design, implementation, and impact. Section \ref{sec:IV-PIN Watermarking} describes the PIN-based Sparse QIM watermarking process as a complementary layer of protection. In Section \ref{sec:V-Results} we present experimental results across various architectures, analyzing the effectiveness of both techniques under different AI configurations, with respect to key lengths in the first stage and PIN lengths in the second. Finally, Section \ref{sec:VI-Conclusion} concludes the study and outlines directions for future work, including the integration of these mechanisms into broader security and licensing frameworks.

\section{Problem Definition and Relevant Work}
\label{sec:II-Problem Definition and Relevant Work}

Despite the enviable level of robustness of widely used and practically proven multimedia protection algorithms, piracy often finds an adequate way to avoid restrictions. As long as the time and cost of breaking protection is less than the cost of creating content, authors will face this problem. The development of AI architecture and training models is very expensive and time-consuming, and expenses could be significantly higher than the costs of multimedia. Unfortunately, this encourages the misuse of artificial models instead of spending time developing and training their own models, which includes, of course, large multimedia training data sets. Therefore, we will not consider the protection whose destroying requires large expenses for architecture modification and/or additional costs of learning. In other words, we aim to find out a solution for the main problem: protecting the intellectual property of already developed and learned artificial models.

Excluding legal, social-engineering and supply-chain attacks, which lie outside the scope and remit of our work, and setting aside the extreme case of complete retraining that, given sufficient data and time, effectively produces a new learned model using only the known architecture, the contemporary threat landscape for watermarks and related protections is characterized by a range of malicious and non-malicious modifications. These include: post-lock fine-tuning, where an adversary continues training the model with new data or via transfer learning; model extraction or distillation, in which a surrogate model is constructed from black-box access (API) to reproduce the victim model’s behavior without its original weights; pruning, sparsification and compression such as weight removal and quantization; weight reinitialization, randomization or permutation attacks that alter weight arrangements; watermark overwriting or forging, where an attacker inserts a new watermark that masks or replaces the original; key-search or brute-force and adaptive search strategies against the PIN; and collusion or ensemble attacks that combine multiple model copies or outputs to remove traceable signals or reproduce behavior without the watermark.

However, the main problem mentioned spreads to a couple of technical issues due to the very simple Neural Network (NN) architecture representation. Roughly speaking, the employment of any artificial model depends on a few lines of code and the multidimensional matrix of weight and bias coefficients, and thus any modification of their values leads to a reduction of accuracy even a drastic degradation of the model's usability. Our idea is to employ exactly that characteristic to manage the usability of the model by an encrypted modification of a multidimensional matrix of coefficients, maintaining the simplicity and speed of the used code.

As expected, following the good results in the multimedia protection area and defined specific industrial requirements, recent work in the field of AI protection has crystallized three general directions of proposed techniques: i) private computing, ii) encryption, and iii) watermarking. 

\subsection{Private Computing}
\label{subsec:II-A-Private Computing}

Emerging technologies in many domains, such as medicine, biomedicine, and finance, use protected NN techniques, which allow employing artificial intelligence to achieve better learned models without the raw model distribution. From the computational learning theory \cite{Bshouty1997} and a basic general idea of distributed computing \cite{Peleg2000}, as an answer to industrial needs arises new approaches to distributed learning \cite{Balcan2012} and later derived techniques as Federated learning \cite{Konecny2015,McMahan2017,Savazzi2020}, SplitNN \cite{Gupta2018}, and Synchronous stochastic gradient descent (SGD) \cite{chen2017}. 
The principle of private computing is the restricted distribution of learning data between clients and a centralized computing resource (server). 

During the learning process, Federated learning and SGD allow sharing only the gradients, activations, and weight updates from the server to clients and vice versa. However, although the distributed information is limited, it discloses the NN architecture and thus the raw structure of weights and biases can be reconstructed by clients. The SplitNN approach additionally restricts the architecture and weights sharing, allowing the client to train up to a cut layer of the network. Then, the server trains the rest of the layers and sends to the client the gradient computed by the whole network backpropagation.

Although the private computing techniques evidently ensure a certain level of NN protection, their main purpose is distributed learning. In other words, the usability of distributed models in mass-use industry that exclude learning is limited, especially given the fact that the architecture and parameters are not fully protected.

\subsection{Encryption}
\label{subsec:II-B-Encyption}

A key strategy for protecting NN models involves enabling computations directly over encrypted data, preserving both input privacy and model confidentiality. CryptoNN \cite{xu2019cryptonn} supports encrypted training and inference using functional encryption to perform secure linear operations. In contrast, CryptoNets \cite{gilad-bachrach2016cryptonets}, an earlier approach, focuses on encrypted inference with Homomorphic Encryption (HE), achieving good performance but lacking training support.

This problem has been approached in scientific literature from different perspectives. The work presented by Bost et al. \cite{Bost2015} extends encrypted inference to traditional classifiers using additive HE, offering generality but no Deep Learning (DL) and Machine Learning (ML) support. CryptoML \cite{Mirhoseini2016} addresses scalability by combining secret sharing with approximate computing, enabling efficient, privacy-preserving training on large datasets through data sketching. SecureML \cite{Mohassel2017secureml} and CryptoDL \cite{Hesamifard2017cryptodl} extend privacy-preserving ML to more complex models, including deep NN. SecureML uses secret sharing and optimized garbled circuits to enable efficient training of models like logistic regression and simple NN. In contrast, CryptoDL focuses on encrypted inference over deep convolutional networks using HE, replacing non-linear activations with low-degree polynomial approximations. The majority of these works demonstrate the growing feasibility of encrypted deep learning with minimal accuracy loss and improved scalability.

In a different direction, Shokri and Shmatikov \cite{Shokri2015} propose a collaborative training framework where multiple parties jointly train a model by selectively sharing gradients instead of raw data. Their distributed protocol supports privacy-preserving training without encryption, relying on asynchronous SGD and optionally differential privacy to reduce information leakage during parameter exchange \cite{Phong2018}. González-Serrano et al. \cite{Gonzalez2018} focus on training classical ML models like logistic regression and SVM using partial HE in a multi-key setup, where each data owner encrypts their inputs independently. They introduced a three-party protocol that relies on a semi-trusted computation provider to perform operations unsupported by the encryption scheme. Their design balances privacy and efficiency, enabling encrypted training with minimal involvement from data owners and without full reliance on fully HE (FHE).

To extend HE techniques to deeper NN, Chabanne et al. \cite{Chabanne2017} propose a privacy-preserving classification method based on FHE. They address the limitations of CryptoNets on deep models by combining low-degree polynomial approximations of rectified linear unit (ReLU) with batch normalization, maintaining stable input distributions. Other frameworks, such as 5Gen \cite{Lewi2016}, laid the groundwork for exploring encrypted neural computing at scale, offering tools and optimizations that informed subsequent developments.

\subsection{Watermarking}
\label{subsec:II-C-Watermarking}

Watermarking techniques have emerged as a complementary approach, embedding ownership signals directly into NN. Uchida et al. \cite{Uchida2017} first proposed embedding watermarks in model parameters through a regularization term during training, ensuring robustness to fine-tuning and pruning. Zhang et al. \cite{Zhang2018} extended this to black-box scenarios by using crafted trigger patterns for remote verification, while Darvish Rouhani et al. \cite{Rouhani2019} further generalized watermarking with DeepSigns, embedding ownership signals into activation distributions, supporting both white- and black-box settings while maintaining robustness against common removal attacks.

More recently, Tang et al. \cite{Tan2023} proposed a dynamic watermarking technique designed to withstand model extraction attacks, embedding watermarks resilient to functional stealing and maintaining high verification accuracy even after partial model duplication. From another point of view, the research of Wang et al. \cite{Wang2023} introduce a plug-and-play watermarking framework that embeds watermarks into NN via a proprietary model (PTYNet) without modifying the target model’s parameters. This approach ensures strong robustness against removal attacks while preserving the original model’s performance and enabling efficient deployment.

Despite the robustness of existing watermarking techniques, Gu and Qian \cite{Gu2023} show that watermarks can be removed through adversarial neuron pruning, a black-box, data-driven method that targets sensitive neurons without needing trigger knowledge. Their approach reduces watermark success rates to below 4\% with minimal accuracy loss, highlighting the vulnerability of current watermarking schemes to model modification attacks. To address such limitations, Wang et al. \cite{Wang2023reversible} introduce a dynamic watermarking method that leverages a reversible image hiding network to embed undetectable watermarks into NN. Their approach enhances robustness against removal while preserving model performance across diverse datasets.

In parallel, the work of Adi et al. \cite{Adi2018} reinterprets the concept of backdoors, typically viewed as a security threat, as a means for black-box watermarking. Their method embeds a trigger set into the model’s decision function, enabling remote ownership verification without requiring white-box access. They further provide a formal cryptographic framework linking watermarking and backdooring, ensuring properties such as unforgeability and unremovability under standard assumptions.

\section{Key-based Permutation Locking Mechanism}
\label{sec:III-Key-based Permutation Locking Mechanism}

Our concept introduces a novel cryptographic-inspired mechanism for AI models protection and weight obfuscation through key-based index reshuffling. In contrast to conventional paradigms where weights are statically stored and accessed directly, this approach encodes the AI’s weight vectors by deterministically permuting their indices based on a key vector, effectively acting as a PIN. The key vector $\mathbf{K}$, composed of $k$ integer-valued elements, is transformed into a permutation operator $\bm{\pi}_{\mathbf{K}}$ \footnote{In combinatorics and group theory, the symbol $\bm{\pi}$ is often used to denote a permutation function over a set. Since our mechanism involves permuting indices of a weight vector, we borrowed this notation from that tradition to emphasize that the function $\bm{\pi}_{\mathbf{K}}$ is a bijective mapping from the index set $\{1,2,\dots,N\}$ onto itself. Therefore, $\bm{\pi}_{\mathbf{K}}(i)$ means: “given key $\mathbf{K}$, which index should position $i$ in the weight vector map to?} that reorders the elements of a weight vector $\mathbf{W}$, yielding a new vector $\mathbf{W}'=\bm{\pi}_{\mathbf{K}}(\mathbf{W})$. This reordered weight configuration represents a locked state of the network: without the correct key, the underlying functional mapping defined by the AI model becomes unintelligible or at least significantly degraded.

Mathematically, the key vector may be resized or hashed to match the dimensionality of the weight vector and then used to generate a permutation through a deterministic function such as sorting. The original computational graph remains unchanged except that all operations depending on weights now access the permuted indices. To restore the original functional behavior, an inverse permutation $\bm{\pi}_{\mathbf{K}}^{-1}$  is applied, reconstructing the original weight arrangement. This mechanism acts orthogonally to standard key-based weight generation methods (e.g., hypernetworks \cite{Ha2016} or attention systems \cite{Vaswani2017}), which typically produce weight values as a function of a key but do not alter the structural positioning of weights. Furthermore, unlike attention mechanisms that leverage soft and continuous selection across memory-like structures, this method imposes a hard, discrete permutation that enhances security properties while introducing minimal computational overhead.

\begin{figure*}[ht]
  \centering
  \includegraphics[width=\textwidth]{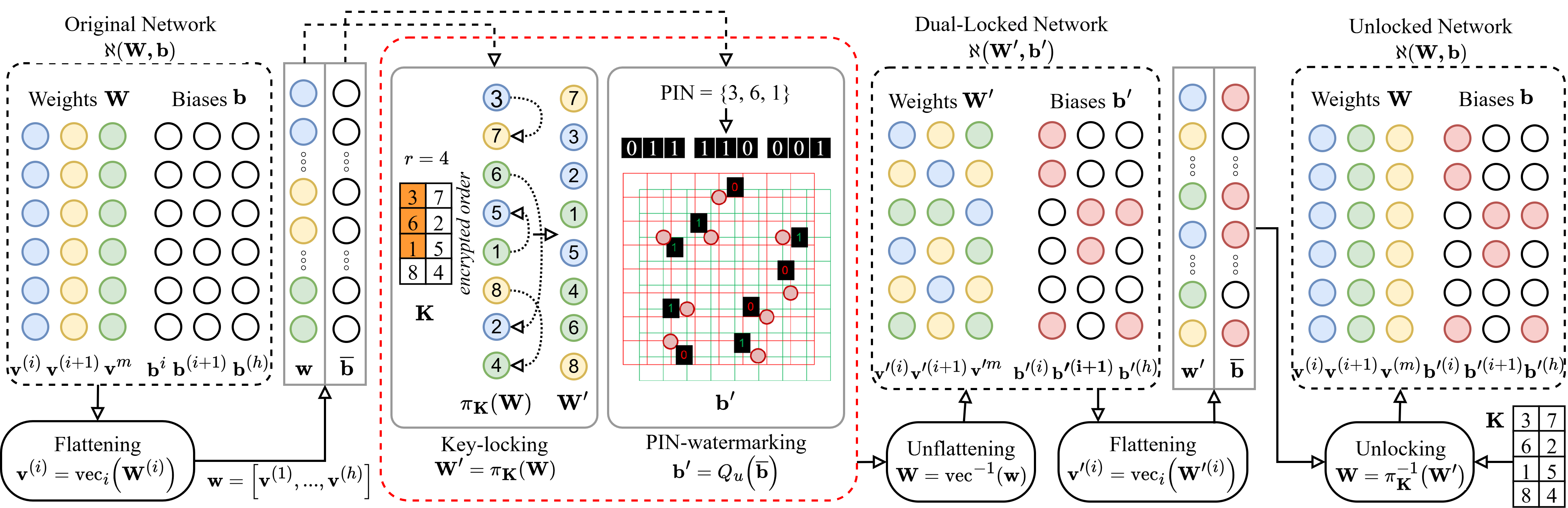}
  \caption{Illustration of the proposed locking and unlocking mechanism, where the function 
    $\aleph(\mathbf{W},\mathbf{b})$ maps the weight vector $\mathbf{W}$ and bias vector 
    $\mathbf{b}$, while the permutation operator $\bm{\pi}_{\mathbf{K}}$ and its inverse 
    $\bm{\pi}_{\mathbf{K}}^{-1}$ are defined in \eqref{Eq.5} and \eqref{Eq.8}, respectively. The central part of the illustration depicts Stage 1, representing the key-based adaptive permutation, and Stage 2, corresponding to the PIN QIM watermarking process.}
  \label{Fig.1.}
\end{figure*}

This framework (Fig. \ref{Fig.1.}.)\footnote{In the illustration, permutation is visually represented by reshuffled colors of the neurons to enhance perceptual clarity, as the actual notation or numerical structure of the permuted weights would not be visually discernible.} may serve as a foundation for neural model locking, secure model distribution, and controlled access to pretrained weights, especially in environments demanding resistance to unauthorized inference or tampering. It positions the neural network as a cryptographic object, whose activation requires possession of a cryptographic key, not to decrypt data, but to reconstruct the very structure of its parameters. 

\subsection{NN Formalism}
\label{subsec:III-A-NN Formalism}

For the NN $\aleph(\mathbf{W},\mathbf{b})$ composed of $h$ layers, each associated with a weight and biases matrices respectively. Let $\mathbf{W}^{(i)} \in \mathbb{R}^{n_i \times n_{i+1}}$
denote the weight matrix corresponding to the pair of $i$-th and $(i+1)$-th layers, where $i=\{1,2,\ldots,h-1\}$, and where $n_i$ and $n_{i+1}$ represent the numbers of neurons in these two consecutive layers, or dimensionality of that weight matrices. The $\mathbf{b}^{(i)} \in \mathbb{R}^{n_i}$
denote the $i$-th layer biases vectors with dimensionality $n_i\times 1$.

\subsection{Flattening and Concatenation}
\label{subsec:III-B-Flattening}

To consolidate all parameters of the network into a single unified representation suitable for reshuffling‐based locking, both the weight matrices and bias vectors are vectorized. Each weight matrix $\mathbf{W}^{(i)}$ is vectorized using the standard column‐wise flattening operator $\mathrm{vec}_{i}(*)$, yielding a vector $\mathbf{v}^{(i)} = \mathrm{vec}_{i}(\mathbf{W}^{(i)}) \in \mathbb{R}^{\,n_{i} \times n_{\,i-1}}$. This transformation is guided by an architecture vector ${\Lambda _\aleph } = \left\{ {{n_0},{n_1},{n_2}, \ldots ,{n_{h - 1}}} \right\}$, which defines a corresponding set of shape‐tuples $S_\mathbf{W} = \{(n_{i},\,n_{\,i-1}) \mid i = 1,2,\dots,h-1\}$ for weights and $S_\mathbf{b} = \{(n_{i},1) \mid i = 1,2,\dots,h-1\}$ for biases. The total flattened weight and bias vectors $\mathbf{w}^{(i)} \in\mathbb{R}^{s_{w}}$ and $\bar{\mathbf{b}} \in \mathbb{R}^{s_{b}}$, respectively, where sum of all number of weights elements $s_{w} = \sum_{i=1}^{h} n_{i}\,n_{\,i-1}$ is then formed by concatenating all individual $\mathbf{v}^{(i)}$ vectors in order:
\begin{equation}
\mathbf{w} = \bigl[\mathbf{v}^{(1)},\,\mathbf{v}^{(2)},\,\dots,\,\mathbf{v}^{(h)}\bigr]^{T}
\label{Eq.1}
\tag{1}
\end{equation}
Similarly, the bias vector $\bar{\mathbf{b}}\in \mathbb{R}^{s_b}$, where $s_b = \sum_{i=1}^{\,h-1} n_i$, is constructed by concatenating all bias vectors:
\begin{equation}
\bar{\mathbf{b}} = \bigl[\mathbf{b}^{(1)},\,\mathbf{b}^{(2)},\,\dots,\,\mathbf{b}^{(h-1)}\bigr]^{T}
\label{Eq.2}
\tag{2}
\end{equation}
\subsection{Unflattening (Reconstruction)}
\label{subsec:III-C-Unflattening}

For subsequent reconstruction, the inverse mapping utilizes the stored dimensionality tuples $(n_i,\,n_{i-1}) \in S_\mathbf{W}$ for each weight segment and $(n_i,\,1) \in S_\mathbf{b}$ for each biases vector. The global weight vector $\mathbf{w}$ from \eqref{Eq.1} is partitioned into contiguous slices corresponding to the original dimensions of each layer. Each slice is then reshaped back into its 2D matrix form:
\begin{equation}
\mathbf{W}^{(i)}= \operatorname{vec}^{-1}\bigl(\mathbf{w}[\,p_{i}:p_{i+1}\,]\bigr)
\label{Eq.3}
\tag{3}
\end{equation}
where $p_{i+1} = p_i + n_i\,n_{i-1}$. Likewise, the bias vector from \eqref{Eq.2} is partitioned by lengths $n_i \in S_\mathbf{b}$ and mapped back as:
\begin{equation}
\mathbf{b}^{(i)}=\bar{\mathbf{b}}\bigl[q_i : p_{i+1}\bigr]
\label{Eq.4}
\tag{4}
\end{equation}
where $q_{i+1} = q_i + n_i$.

\subsection{Locking the AI-Index Reshuffling}
\label{subsec:III-D-Locking}

Assume $\mathbf{K} = [\,k_1,\,k_2,\,\dots,\,k_r]$, $\mathbf{K}\in \mathbb{N}^r |r \leqslant \mathbf{s_w}$, is the key vector (or PIN) in our notation that defines a permutation $\bm{\pi}_{\mathbf{K}} : \{1,2,\dots,\mathbf{s_w}\} \to \{1,2,\dots,\mathbf{s_w}\}$. Then the reshuffled weight vector is $\mathbf{W}' = \bm{\pi}_{\mathbf{K}}(\mathbf{W})$. To perform basic reshuffling, we simply hash $\mathbf{K}$ to length $\mathbf{s_w}\mathbf{:}\overline{\mathbf{K}} \in \mathbb{N}^{\mathbf{s_w}}$, and then compute a pseudorandom permutation $\bm{\pi_\mathbf{K}}$ as the sort indices of $\mathbf{\overline{K}}$: 
\begin{equation}
\bm{\pi}_\mathbf{K}(i) = \operatorname*{arg}sort\bigl(\overline{\mathbf{K}}\bigr)[i], 
\quad i = 1,\,2,\,\dots,\,\mathbf{s_w}.
\label{Eq.5}
\tag{5}
\end{equation}
Thus, applying the permutation to the $i$-th weight yields $\mathbf{w}_i = \mathbf{w}_{\bm{\pi_\mathbf{K}}(i)}$, and the locked weight vector becomes
\begin{equation}
\mathbf{W}'=\bm{\pi}_\mathbf{K}(\mathbf{W})
=\bigl[\,
\mathbf{{w}_{\bm{\pi}_K}}(1),\,
\mathbf{{w}_{\bm{\pi}_K}}(2),\,
\dots,
\mathbf{{w}_{\bm{\pi}_K}}(\mathbf{s_w})
\bigr].
\label{Eq.6}
\tag{6}
\end{equation}
Similarly, for the bias vector (with $\mathbf{K} \in \mathbb{N}^r|r \le \mathbf{s_b}$, $\mathbf{\bm{\pi}_{K} :} \{1,2,\dots,\mathbf{s_b}\} \to \{1,2,\dots,\mathbf{s_b}\}$, and
\begin{equation}
\mathbf{b}'=\bm{\pi}_\mathbf{K}(\mathbf{b})
=\bigl[\,
\mathbf{{b}_{\bm{\pi}_K}}(1),\,
\mathbf{{b}_{\bm{\pi}_K}}(2),\,
\dots,\,
\mathbf{{b}_{\bm{\pi}_K}}(\mathbf{s_b})
\,\bigr]
\label{Eq.7}
\tag{7}
\end{equation}
with $\mathbf{\bm{\pi}_{K}}(i) =\operatorname*{arg}sort\bigl(\overline{\mathbf{K}}\bigr)[i], i = 1,2,\dots,\mathbf{s_b}$.

\subsection{Locking the AI-Adaptive Index Permutation}
\label{subsec:III-E-Locking Adaptive}

The core of our framework is a cryptographic-like, key-guided adaptive index selection and permutation mechanism. Let $\mathbf{w}$ from \eqref{Eq.1} be the flattened vector containing all weights of the NN, as defined by the concatenation of $\mathbf{v}^{(i)} = \text{vec}_i\left(\bm{W}^{i}\right) \in \mathbb{R}^{n_i \times n_{i-1}}$, over all layers. We define a descending-order index vector
\begin{equation}
\mathbf{L} =  \operatorname*{arg}sort(-|\mathbf{w}|)
\tag{8}
\label{Eq.8}
\end{equation}
Here, $\mathbf{L} = [l_1, l_2, \dots, l_\mathbf{s_w}]$, where $|\mathbf{w}_{l_1}| \geq |\mathbf{w}_{l_2}| \geq \dots \geq |\mathbf{w}_{l_\mathbf{s_w}}|$. Let $r \in \mathbb{N}^\mathbf{s_w}$ be a pre-selected key length parameter. Based on $\mathbf{L}$, we define two index subsets: $\mathbf{L}_{hi} = [l_1, l_2, \dots, l_r]$ and $\mathbf{L}_{lo} = \left[ l_{\mathbf{s_w} - r + 1}, l_{\mathbf{s_w} - r + 2}, \dots, l_{\mathbf{s_w}} \right]$ the bottom-$r$ least significant weights. To incorporate cryptographic randomness into the locking mechanism, we independently apply a uniform random permutation to every row of the adaptively selected index vectors: $\mathbf{K}^{(1)}_{lock} = \operatorname*{shuffle}(\mathbf{L}_{hi})$ and $\mathbf{K}^{(2)}_{lock} = \operatorname*{shuffle}(\mathbf{L}_{lo})$. Thus, the final key hash matrix is:
\begin{equation}
\mathbf{K} = 
\begin{bmatrix}
\mathbf{K}^{(1)}_{lock} \\
\mathbf{K}^{(2)}_{lock}
\end{bmatrix}
\in \mathbb{N}^{2 \times r}
\tag{9}
\label{Eq.9}
\end{equation}
This key $\mathbf{K}$ identifies $r$ pairs of positions whose weight values will be swapped during the permutation process, forming a lightweight encryption-like mechanism that ``locks'' the network by altering the positions of high- and low-magnitude weights.

\subsection{Inverse Permutation - Unlocking AI model with the PIN}
\label{subsec:III-F-Inverse Permutation}

To restore the model’s functionality, we will simply apply the general inverse permutation $\bm{\pi}_{\mathbf{K}}^{-1}$, satisfying:
\begin{equation}
\bm{\pi}_{\mathbf{K}}^{-1}\left( \bm{\pi}_{\mathbf{K}}(i) \right) = i
\tag{10}
\label{Eq.10}
\end{equation}
Then, the original weights and biases are recovered as:
\begin{equation}
\mathbf{W} = \bm{\pi}_{\mathbf{K}}^{-1}\left( \mathbf{W}' \right).
\tag{11}
\label{Eq.11}
\end{equation}
After recovery of all original weight and biases vectors, reconstruction of the model parameters proceeds by reshaping into the original set of matrices $\left\{ \mathbf{W}^{(i)} \right\}$ using the stored shape information $S_\mathbf{W}$, as described previously. If biases were not permuted, they are appended unchanged. Otherwise, the same index-swapping inversion applies to the bias vector $\left\{ \mathbf{b}^{(i)} \right\}$ using a separate key with shape information $S_\mathbf{b}$, and the same logic holds.

\section{PIN Watermarking}
\label{sec:IV-PIN Watermarking}

One important issue that our algorithm does not currently address is the unrestricted distribution and resale of the locked AI model along with the generated key. Specifically, the key owner can freely distribute an unlimited number of copies of the locked network using the same key. To prevent this and make the PIN key truly personalized, we propose binding key-specific information to the model itself. The idea is to embed a watermark into the weights or bias coefficients that corresponds to the key, or to a part of the key, depending on the length of the bias vector and the network architecture.

To embed the watermark, we will use Quantized Index Modulation (QIM) \cite{Chen2001}, which we have previously applied successfully for protecting three-dimensional models. Unlike watermarking techniques adapted to vertex structures in 3D geometry~\cite{Vasic2013ldpc}, in this context, we operate on a one-dimensional flattened vector of bias coefficients $\bar{\mathbf{b}}$ \eqref{Eq.2} or a flattened vector of weight coefficients $\mathbf{w}$ \eqref{Eq.1}. As a result, there is no need for coordinate transformation into spherical or cylindrical domains. Instead, the embedding is performed directly in Euclidean space using quantized discrete values.

Let $\mathbf{u}\in\{0,1\}^{m}$ denote the binary watermark sequence and $\mathbf{x} \in \mathbb{R}^{m}$ the cover sequence, which in our case corresponds to the selected bias or weight coefficients of the locked neural network, with a length of both vectors $m$. The embedding process produces a watermarked vector $\mathbf{y}\in\mathbb{R}^{m}$, where the displacement $\bm{\omega} = \mathbf{y} - \mathbf{x}$ represents the watermark signal. The distortion introduced by watermarking must be bounded, $d(\mathbf{x},\mathbf{y})< mD$, where $D$ is the maximum allowed distortion per component. While the distortion measure $d(*,*)$ can be the Euclidean ($\ell_2$) norm, it may also adopt more perceptually aligned measures such as Hausdorff distance, though for our PIN-protected NN architecture, Euclidean distortion suffices.
\begin{equation}
d_H(\mathbf{x},\mathbf{y})=\max\left\{
\mathop{\sup}\limits_{\mathbf{x}\in\mathbf{X}}
\mathop{\inf}\limits_{\mathbf{y}\in\mathbf{Y}}
d(\mathbf{x},\mathbf{y}),
\mathop{\sup}\limits_{\mathbf{y}\in\mathbf{Y}}
\mathop{\inf}\limits_{\mathbf{x}\in\mathbf{X}}
d(\mathbf{x},\mathbf{y})
\right\} 
\tag{12}
\label{Eq.12}
\end{equation}
The QIM operates independently on the corresponding components $u_i\in \{0,1\}$ and $x_i\in \mathbb{R}$ from $\mathbf{u}$ and $\mathbf{x}$. The embedding is based on two scalar quantizers uniform quantizers $\bm{Q}_0$ and $\bm{Q}_1$, defined as:
\begin{equation}
\bm{Q}_{u}(x)=\Delta\left\lfloor \frac{1}{\Delta} \left( x-(-1)^u\frac{\Delta}{4}\right)\right\rfloor + (-1)^u \frac{\Delta}{4}
\tag{13}
\label{Eq.13}
\end{equation}
where $\Delta$ is the quantization step and $\lfloor * \rfloor$ denotes rounding to the nearest integer. This can also be interpreted as dithering the quantization level by $\pm \Delta/4$ based on the watermark bit $u$. The resulting watermarked signal is:
\begin{equation}
y_i = \bm{Q}_{u_i}(x_i), \quad \forall i = 1, \ldots, m
\tag{14}
\label{Eq.14}
\end{equation}
The minimum embedding error is $\Delta/2$, and assuming uniform error distribution in $[-\Delta/2, \Delta/2]$, the mean squared error (MSE) is $\Delta^2/12$. To minimize functional distortion when embedding the watermark in any sensitive $\bar{\mathbf{b}}$ or $\mathbf{w}$ vector, we employ Sparse QIM. Rather than embedding each watermark bit in a single coefficient, the bit is spread across an $T$-length segment of the cover vector $\mathbf{x}_T\in\mathbb{R}^T$ for $T \leq \mathbf{s_b}$ or $T \leq \mathbf{s_w}$ applying to weights or biases vector respectively. A fixed projection vector $\bm{\varepsilon} \in \mathbb{R}^T$, $\| \bm{\varepsilon} \| = 1$, is used to project $\mathbf{x}_T$ onto a 1D subspace. The scalar projection $\alpha = \langle \mathbf{x}_T, \bm{\varepsilon} \rangle$ is quantized using QIM:
\begin{equation}
\alpha'=\bm{Q}_{u}(\alpha),\quad \mathbf{y}_T=\mathbf{x}_T + (\alpha' - \alpha) \bm{\varepsilon}
\tag{15}
\label{Eq.15}
\end{equation}
At detection, the watermarked segment $\mathbf{r}_T$ is projected onto the same direction $\bm{\varepsilon}$, and the embedded bit is recovered as:
\begin{equation}
\hat{u} = \arg \min_{u \in \{0,1\}} \left\| \mathbf{r}_T \bm{\varepsilon}_T - \bm{Q}_{u}(\mathbf{r}_T \bm{\varepsilon}_T) \right\|_2
\tag{16}
\label{Eq.16}
\end{equation}
The robustness of Sparse QIM increases with $T$, offering better resilience to noise and tampering. In the context of PIN-based AI locking where bit embedding must minimize performance loss, the projection direction can be designed using knowledge of gradient sensitivity to minimize performance loss in the unlocked model.

\section{Results}
\label{sec:V-Results}
To experimentally validate the proposed AI locking mechanism, we developed a Python-based software framework \cite{Vasic_Software} implementing the workflow described in Sections \ref{sec:III-Key-based Permutation Locking Mechanism} and \ref{sec:IV-PIN Watermarking}. The framework supports the design, training, and evaluation of fully connected and convolutional networks, incorporating our key-driven index reshuffling module for controlled permutation of weight coefficients. This enables systematic assessment of the influence of key parameters, such as scope of application and architectural complexity, on network performance. The following subsections present results obtained on fully connected and convolutional models, while transformer architectures are analyzed separately.

Using this framework to validate our assumptions and analyze the behavior of NNs under the proposed algorithm, we designed three NNs with different configurations, employing the MNIST dataset \cite{Lecun2010} for training, validation, and testing. In addition, we incorporated standard ResNet architectures (ResNet18, ResNet20, and ResNet32) \cite{He2016} evaluated on the ImageNet-1K \cite{Deng2009ImageNet} and CIFAR-10/100 \cite{Krizhevsky2009LearningML} datasets to extend the experimental validation to convolutional models. The representation of the CNN networks was revised to show the dimensions of the classification fc.weights layer and last convolutional weights layer from the Torch state-dict representation, which for ResNet18, for example, contains 41 weight tensors and 21 bias tensors of varying dimensions, along with tensor scalars and running statistics. The tested networks are described by the following architecture vectors $\Lambda_{\aleph}$:

$
\begin{aligned}
\Lambda_{\aleph_1} &= \{784, 100, 30, 10\}, \\
\Lambda_{\aleph_2} &= \{784, 100, 90, 80, 70, 60, 50, 40, 30, 10\}, \\
\Lambda_{\aleph_3} &= \{784, 100, 50, 50, 30, 10\},\\
\Lambda_{\aleph_4} &= \{(1000,512)...(512,512,3,3)\},\\
\Lambda_{\aleph_5} &= \{(10,64)...(64,64,3,3)\},\\
\Lambda_{\aleph_6} &= \{(100,64)...(64,64,3,3)\},\\
\end{aligned}
$

After training on a dataset of $70,000$ labeled images and validating on an additional $20,000$ samples, NN1, NN2 and NN3 were evaluated on a separate test set of $10,000$ samples, achieving accuracies of $96.48\%$, $96.58\%$, $93.29\%$, $95.04\%$, and $96.21\%$, respectively. The CNN networks (NN4, NN5 and NN6) of different architectural complexities, under a quick test (first $4$ shards and a maximum of $20$ batches) using the ImageNet-1K and CIFAR-10/100 validation sets, achieved initial accuracies of $70.59\%$, $92.12\%$, and $70.14\%$, respectively. In this study, specific training parameters, such as the choice of activation function, number of epochs, mini-batch sizes, and learning rates, are considered secondary, as the primary focus is on the proposed algorithm.

Nevertheless, the observed differences in accuracy, even among networks with similar architectures, may reflect training-related variance. Our initial objective was to design and train fully connected NNs with varying architectures, differing in the number of hidden layers and the number of neurons per layer, while for CNNs and transformer models we relied on pretrained architectures. This allowed us to assess the effectiveness of our algorithm relative to model architecture and the total number of weight and bias parameters.

The primary objective is to achieve significant degradation of the model's accuracy by permuting the weight coefficients according to a given key defined as a set of index pairs. Since the proposed method does not involve altering the coefficient values, introducing any watermark-based value sequences, or applying quantization, it is logically expected that the degradation in accuracy will solely depend on the length of the applied key. However, in addition to confirming this expectation, the results also revealed several other interesting phenomena, which will be discussed sequentially in the following sub-sections.

\subsection{Locking the NN-Index Reshuffling}
\label{subsec:V-A-Locking the NN-Index Reshuffling}

By applying the neural network locking algorithm described in the previous Section \ref{sec:III-Key-based Permutation Locking Mechanism}, we first tested networks defined by \eqref{Eq.6} and locked by reshufelling operator from \eqref{Eq.5} using randomly generated keys of various lengths $r = \left\{{10},{{10}^2},{{10}^3},{{10}^4},{{10}^5} \right\}$, verifying the dimension of the flattened weight coefficient vector \eqref{Eq.1} and automatically correcting the key length for cases where $r \geqslant h$  to $r = h$. As a result, it is important to note that the reported results may include inconsistent, adjusted values of networks with smaller overall dimensions, in terms of the total number of constituent neurons (for example, $81,700$ and $87,700$ for NN1 and NN3), which does not apply to CNN and transformer architectures due to the incomparably larger number of weights in the models’ tensors.

In the case of random reshuffling, the locking effect was generally observed only for large key lengths, as anticipated during the design of our algorithm, since degrading accuracy to around $10\%$ requires permuting a substantial portion of the weight coefficients. However, we also observed unexpected anomalies, where certain larger networks exhibited an increase in accuracy when reshuffling was applied with shorter key lengths. While we do not yet have a precise explanation for this phenomenon, one possible reason is that the baseline network accuracy was not sufficiently high prior to locking. These results highlight the limited effectiveness of random reshuffling as a standalone approach and motivated our transition toward adaptive index permutation. For completeness, the illustration is shown in the Appendix (Fig. \ref{Fig.5.}), while detailed tables are omitted for brevity.

\begin{figure}[H]
\centerline{\includegraphics[width=18.5pc]{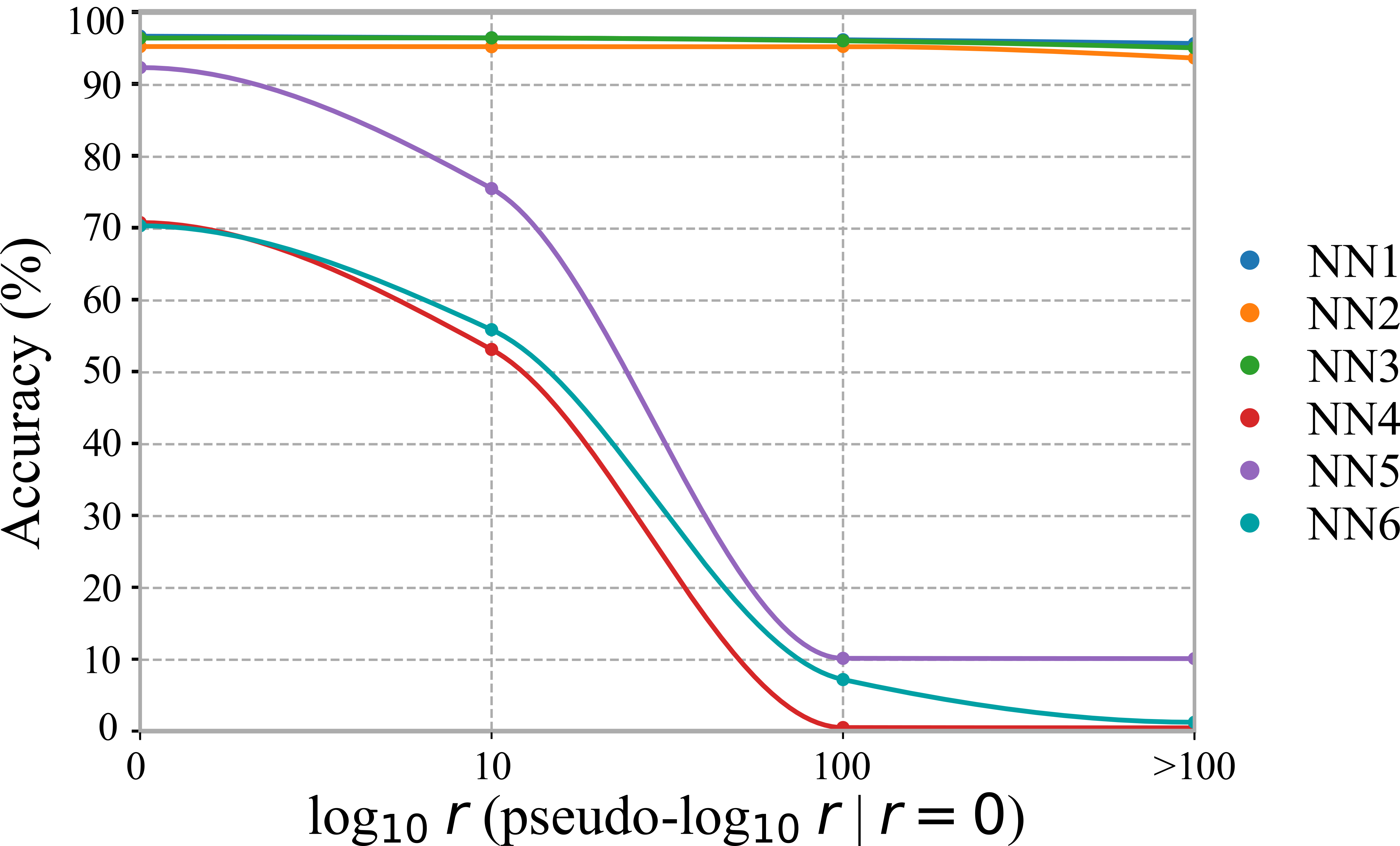}}
\caption{Accuracy degradation curves for all neural networks (NN1--NN6) as a function of key length $r$ under random reshuffling of bias indices \eqref{Eq.7}.}
\label{Fig.2.}
\end{figure}

Surprisingly poor results in terms of accuracy degradation were observed when applying the random permutation of bias coefficient \eqref{Eq.7} indices in fully connected networks, while convolutional architectures exhibited a markedly different behavior, as illustrated in (Fig. \ref{Fig.2.}.).

In fully connected networks, altering the arrangement of bias coefficients had minimal or even paradoxically beneficial impact on accuracy, a phenomenon not observed in convolutional architectures. Although this discourages bias-based index permutation as a locking strategy, the stable behavior under small perturbations supports the validity of PIN watermarking on bias vectors, where short PIN lengths and quantization-based modulation cause negligible distortion.

After this brief digression on the bias locking mechanism, whose detailed results are provided in a Table \ref{Table5} in the Appendix, we return to the discussion of the network locking method based on permutation of weight coefficient positions, which has proven to be more suitable and effective, albeit primarily for large key lengths. A positive aspect of using high-dimensional keys is their robustness, however, our primary interest lies in the construction of a key that yields the best performance even at smaller key lengths.

\subsection{Locking the NN – Adaptive Index Permutation}
\label{subsec:V-B-Locking Adaptive}

This idea of locking a neural network by permuting a specifically selected subset of coefficients is based on the assumption that not all parameters contribute equally to the network’s performance. Therefore, reshuffling the indices of the most influential coefficients from the set $\mathbf{W}$, tailored to the structure of each individual network $\aleph$, is expected to achieve greater accuracy degradation even with shorter key lengths. This subset of important weights is selected based on the fact that coefficients with large absolute values propagate neuron activations to the next layer with minimal attenuation. This logically leads to the assumption that swapping the largest positive and the largest negative absolute values will most significantly compromise the network’s accuracy. The results have confirmed exactly that.

For this part of the locking mechanism, our algorithm was adapted to identify the maximum and minimum values by sorting the index vector $\mathbf{L}$ in descending order according to the values of the flattened weight vector $\mathbf{w}$ \eqref{Eq.1} across all weights of the neural network, as shown in \eqref{Eq.8}. By selecting two subsets of size $r$ from the beginning and end of $\mathbf{L}$, we form a key consisting of index pairs corresponding to the highest and lowest weight coefficient values.

Naturally, such a deterministic locking method would be inherently fragile, as the pattern of key selection is easily predictable. To mitigate this and enhance security, we incorporate cryptographic randomness into the locking mechanism by applying a uniform random permutation over both subsets of high- and low-value indices prior to pairing. This randomized pairing disrupts any obvious structure in the selection process, thereby significantly increasing the unpredictability of the key $\mathbf{K}$ \eqref{Eq.9} and strengthening the resilience of the network locking scheme against reverse-engineering or brute-force attacks.

\begin{table}[H]
\centering
\caption{Accuracy of adaptive locked Neural Networks}
\label{Table1}
\begin{tabular}{>{\raggedright\arraybackslash}p{0.2\linewidth}>{\centering\arraybackslash}p{0.07\linewidth}>{\centering\arraybackslash}p{0.07\linewidth}>{\centering\arraybackslash}p{0.07\linewidth}>{\centering\arraybackslash}p{0.07\linewidth}>{\centering\arraybackslash}p{0.07\linewidth}>{\centering\arraybackslash}p{0.07\linewidth}}
\hline\hline
{NN Name} & \multicolumn{6}{c}{Accuracy as a function of key length}\\
 & {Original} & {4} & {10} & {100} & {1000} & {10000} \\
\hline
NNetwork1 & 96.48\% & 74.29\% & 50.65\% & 9.89\%  & 8.95\%  & 8.14\%  \\
NNetwork2 & 95.04\% & 94.92\% & 94.61\% & 84.99\% & 16.16\% & 9.53\%  \\
NNetwork3 & 96.21\% & 95.80\% & 94.96\% & 38.64\% & 11.45\% & 10.79\% \\
ResNet18 & 70.59\% & 0.93\% & 0.45\% & 0.03\% & 0.19\% & 0.00\% \\
ResNet20 & 92.12\% & 53.03\% & 30.80\% & 9.80\% & 9.88\% & 10.00\% \\
ResNet32 & 70.14\% & 4.78\% & 3.45\% & 0.74\% & 1.00\% & 1.00\% \\
\hline\hline
\vspace{3mm}
\end{tabular}
\noindent\parbox{\linewidth}{\footnotesize Accuracy of neural networks after applying the adaptive locking mechanism with different key lengths $r$. Each entry represents the test accuracy for a specific network configuration and corresponding key size.}
\end{table}

In this experiment, we generated keys of length $r = \left\{ {4,{{10}^1},{{10}^2},{{10}^3},{{10}^4}} \right\}$, adapted to each tested neural network (NN1–NN6), and the corresponding results of accuracy degradation are presented in Table \ref{Table1}.

From the table, we can observe a clear degradation in recognition accuracy of the locked NN1 even with a key of length 100, while its accuracy drops by half when using a key of length only 10. For the other networks, varying degrees of degradation are noticeable, primarily due to their larger sizes, i.e., higher total number of neurons in their architectures and consequently significantly more weight coefficients. In particular, and contrary to the general trend observed above, NN6 shows a dramatic accuracy drop to $4.78\%$ with a key length of $4$, while NN4 decreases even further to $0.93\%$ for the same key length. These unexpected results indicate an exceptionally strong protection effect, with the mechanism proving particularly effective for CNN models and further supporting the validity of this stage of the protection scheme.

\begin{figure}[H]
\centerline{\includegraphics[width=18.5pc]{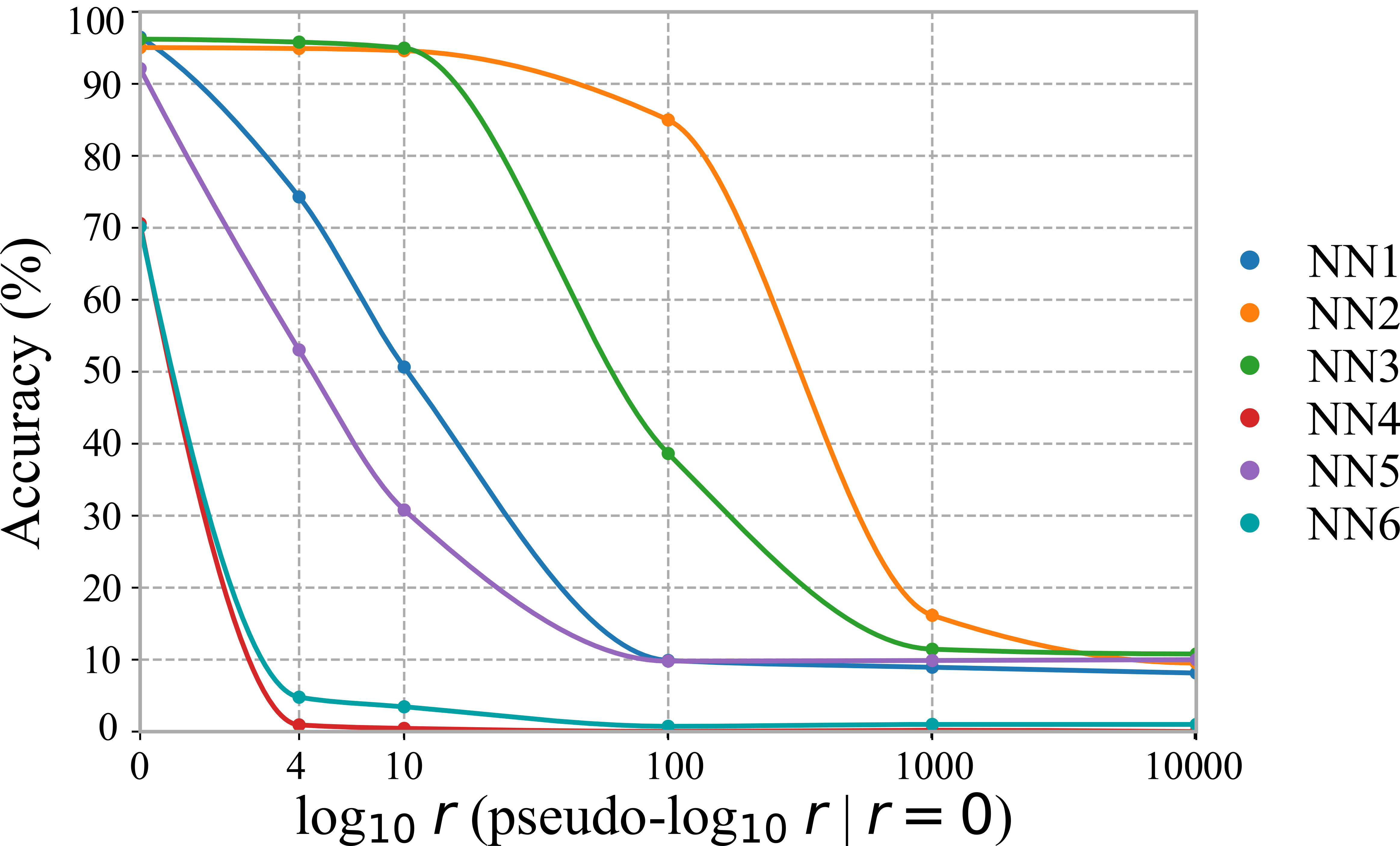}}
\caption{Accuracy degradation curves for all neural networks (NN1–NN6) as a function of key length $r$ \eqref{Eq.9} under adaptive permutation of weight indices in descending-order index vector $\mathbf{L}$ defined in \eqref{Eq.8}.}
\label{Fig.3.}
\end{figure}

However, by carefully examining the second column, it becomes evident that a higher number of neurons does not necessarily yield higher accuracy. On the contrary, the original accuracy of neural network NN2, which possesses approximately 110,900 weight coefficients, is lower than that of NN1 and NN3, which contain only $81,700$ and $87,700$ coefficients, respectively. This leads to the conclusion that well-optimized networks with higher baseline accuracy can be effectively locked using shorter keys. On the other hand, for the trained CNN models, the baseline accuracy (e.g., $70.59\%$ and $92.12\%$ for NN4 and NN5) primarily depends on the dimension of the classification layer ($1000$ and $10$, respectively) and therefore should not be interpreted in relation to the overall number of weight coefficients, which is drastically larger than in the NNs ($11,683,712$ and $271,680$ for the respective models). The general trend and behavior of the networks are illustrated in the figure (Fig. \ref{Fig.3.}).

It can also be observed from the graph that the accuracy degradation curves tend to saturate, technically speaking, for key lengths exceeding $10^4$. This saturation effect was not present when applying the random shuffling method. The key lengths required to effectively disable the networks are an order of magnitude smaller when using the adaptive locking method. Furthermore, for optimized networks, the required key length can be up to two orders of magnitude shorter.

\subsection{Neural Network Unlocking via Key Vector}
\label{subsec:V-C-NN Unlocking}

Since the locking mechanism always uses keys that function as a form of hash table to permute pairs of values, without altering the actual values of the weight coefficients, their restoration to original positions within the architecture vector is reliable and unambiguous through the application of the inverse function defined in \eqref{Eq.10}. This ensures that the unlocked and reconstructed network \eqref{Eq.11} is identical to the original, without the need for additional measurement or verification, making this method highly efficient and elegantly simple.

It should be noted that in the experiments we encountered certain critical cases arising from the more complex structure of transformer models and the procedure of loading them for the application of our protection mechanism and subsequent re-encoding. These issues typically occurred due to incomplete model delivery or mismatched checkpoints during saving. Nevertheless, since our protection mechanism operates at the fundamental architectural level, once the model is correctly loaded, a reliable procedure of dual-locking and unlocking is consistently achieved. These observations further confirm the robustness of our method across diverse architectures.

\subsection{PIN Watermarking}
\label{subsec:V-D-PIN Watermarking}

In order to achieve double-locking of the neural network and thereby protect the owner's intellectual property rights from unauthorized distribution or misuse, this part of the experiment involved watermarking our neural networks using the QIM and Sparse QIM algorithms applied to the biases and weights coefficients. In fact, the previously demonstrated negligible impact of bias coefficient index reshuffling on network accuracy (Tables \ref{Table5} and \ref{Table6}) led us to conclude that watermarking the bias vectors offers the most secure solution — a conclusion that was confirmed by the results.

At this stage, our software first selects the required number of initial key values $\mathbf{K}$ \footnote{The experiment could have used a lightweight cryptographic subkey to select PIN elements, rather than starting from the initial key values. However, the initial elements are easily visible within the long key vector, making it much simpler to perceptually verify the alignment between the selected and extracted PIN vectors} ($u_i$ and $\hat u_i$, $i = 1,\,2,\,\dots,l_\mathrm{PIN}$) after the extraction process \eqref{Eq.16}, based on a user-defined PIN length $l_\mathrm{PIN}$,  while ensuring that the binary equivalent length $l_{binary}$ of the selected PIN index values does not exceed the length $m$ of the host bias coefficient vector ${\mathbf{b}= \mathbf{x}}\in{\mathbb{R}^m}$. This constraint is critical because the maximal dimension of the key matrix $r$ is limited by the length of the flattened weight vector, which always exceeds the length of the flattened bias vector. We address this by first determining the bit length $l_{bit}$ of the largest index selected in the PIN and then enforcing the condition $l_{bit}l_\mathrm{PIN}\leqslant m$.

After verifying that all our requirements are met, the new software function embeds the binary PIN-determined watermark into the values of the biases coefficients \eqref{Eq.13} and/or \eqref{Eq.15}, generating a watermarked vector $\mathbf{y}= \hat{\mathbf{b}}$  that replaces the original biases vector within the network architecture tuple. The modified network is then subjected to the locking procedure described in Section  \ref{subsec:III-E-Locking Adaptive}.

The following Table \ref{Table2} presents the accuracy results of our six experimental networks after watermarking. A clearly negligible impact of dithering and the degradation of biases coefficient values through modulation of quantization indices on the accuracy of the evaluated networks is observed. A noticeable difference is also visible in the measurements for CNN architectures, where biases are associated with each convolutional channel; since the channels carry multidimensional spatial representations, even a small change in bias uniformly affects the entire feature map of that channel. Nevertheless, the deviations in accuracy are minor enough to be comparable to the fluctuations caused by the random selection of test samples from the overall training and validation sets.
\begin{table}[H]
\centering
\caption{Accuracy of watermarked Neural Networks}
\label{Table2}
\begin{tabular}{>{\raggedright\arraybackslash}p{0.2\linewidth}>{\centering\arraybackslash}p{0.10\linewidth}>{\centering\arraybackslash}p{0.10\linewidth}>{\centering\arraybackslash}p{0.10\linewidth}>{\centering\arraybackslash}p{0.10\linewidth}}
\hline\hline
{NN Name} & \multicolumn{4}{c}{Accuracy as a function of PIN Length}\\
 & Original & 4 & 6 & 8 \\
\hline\hline
NNetwork1 & 96.48\% & 96.46\% & 96.47\% & 96.47\% \\ 
NNetwork2 & 95.04\% & 95.03\% & 95.07\% & 95.03\% \\ 
NNetwork3 & 96.21\% & 96.20\% & 96.23\% & 96.21\% \\
ResNet18 & 70.59\% & 68.83\% & 68.93\% & 68.74\% \\
ResNet20 & 92.12\% & 91.31\% & 90.82\% & 90.52\% \\
ResNet32 & 70.14\% & 69.53\% & 69.17\% & 69.20\% \\
\hline\hline
\vspace{3mm}
\end{tabular}
\noindent\parbox{\linewidth}{\footnotesize Accuracy of neural networks after applying the QIM watermarking. Each entry represents the test accuracy for a specific network configuration and corresponding PIN size.}
\end{table}

Moreover, for certain PIN lengths, the accuracy even slightly increased after watermarking (e.g., NN4 and NN5 for a PIN length of $6$). This, along with the fact that PIN length has no significant effect on the accuracy of watermarked networks, supports our assumptions and confirms the validity of both our approaches: i) the key-based permutation locking mechanism and ii) PIN-based watermarking. It also justifies the selection of hosts in both processes, weights indices and biases coefficients, respectively. 

\subsection{Dual-locking Transformers}
\label{subsec:V-E-Duaal-locking Transformers}

Following the evaluation of fully connected and convolutional architectures, the next step was to examine the applicability of the proposed dual-locking mechanism to transformer-based large language models (LLMs). An important aspect of successfully adapting the method to transformers lies in their tensor-centric architecture, which varies substantially across different implementations. To capture this diversity, four representative models were selected for experimentation (ALBERT \cite{Lan2020ALBERT}, BERT-Tiny \cite{Devlin2019BERT}, DistilBERT \cite{Sanh2019distilBERT}, and MiniLM \cite{Wang2020MiniLM}), each pretrained and subsequently fine-tuned on the SST-2 \cite{Socher2013Recursive} dataset. Our dual-locking Python framework was accordingly extended to support transformer variants with sentence-level binary classification heads, ensuring structural compatibility of the locking and unlocking procedures within the transformer computation graph and enabling consistent, comparable evaluation results across all examined architectures.

The large number of parameters that transformers typically operate with represents a limitation for our mechanism, as it necessitates the use of longer keys to achieve effective locking. However, this same property, combined with the highly distributed functional representation across layers, enables us to focus exclusively on the embedding tensors, which exert the strongest influence on model decision behavior and, at the same time, represent the most fragile and therefore the most suitable component for the locking process.

The functional role of embedding layers directly influenced the correction of the first stage of the algorithm, where a key-based permutation of entire vectors associated with token sets was introduced. In practical terms, expressed in NumPy array terminology, we performed a column-wise permutation that effectively disrupted the transformer’s output coherence. The tested transformer models, when locked across individual or all embedding tensors, exhibited the expected consistency of behavior, while the positional embedding tensor was disqualified due to its smaller dimensionality and the consequently limited key length. The obtained results for word embedding locking scenario are summarized in the following Table \ref{Table3}.

\begin{table}[H]
\centering
\caption{Accuracy of watermarked LLMs}
\label{Table3}
\begin{tabular}{>{\raggedright\arraybackslash}p{0.16\linewidth}>{\centering\arraybackslash}p{0.10\linewidth}>{\centering\arraybackslash}p{0.10\linewidth}>{\centering\arraybackslash}p{0.10\linewidth}>{\centering\arraybackslash}p{0.10\linewidth}}
\hline\hline
{LLM} & \multicolumn{4}{c}{Watermarked Accuracy for different key lengths}\\
 & Original & 1000 & 10000 & 15000 \\
\hline\hline
ALBERT      & 92.66 & 88.88 & 56.77 & 50.11 \\ 
BERT-Tiny   & 70.59 & 59.06 & 63.65 & 48.97 \\ 
DistilBERT  & 91.06 & 85.89 & 60.21 & 49.08 \\
MiniLM      & 90.14 & 86.70 & 59.63 & 50.23 \\
\hline\hline
\vspace{3mm}
\end{tabular}
\noindent\parbox{\linewidth}{\footnotesize The binary decision for a specific transformers architecture after applying the adaptive locking mechanism with different key lengths $r$. The last column corresponds to a key vector length of $r =15000$, which satisfies the condition where $N\approx30000$ denotes the total number of available embeddings weights, i.e., the effective channel capacity of the embedding tensor.}
\end{table}

The results in Tables \ref{Table3} and \ref{Table7} show a consistent degradation of transformer accuracy with increasing key length $r$, confirming the effectiveness of embedding-level locking. The observed lower bound of approximately $50\%$ does not reflect a limitation of the proposed mechanism but rather the statistical floor imposed by the binary nature of the SST-2 validation task. Once the embedding space is fully disrupted, model predictions converge to random binary decisions, yielding an accuracy near $0.5$, which effectively represents complete functional locking of the model. Although accuracy values in transformer benchmarks cannot drop below this threshold as in NN or CNN tests, a proper normalization with respect to the random-guess baseline allows direct comparison of degradation levels across all architectures.

However, more important than normalization is the observation that adaptive key formation has limited relevance in this context, since the parameters involved in transformer models differ in their intrinsic functional significance, unlike the more uniformly connected structures of NN and CNN architectures where permutation is applied. Despite this distinction, our method effectively locks all tested models when the key length is chosen proportionally to the architectural scale, while the near-instantaneous execution of both locking stages allows the key length to be set without restriction, limited only by the available channel capacity (See caption of Table \ref{Table3}). The formal definition of parameter “importance” within transformer embedding spaces is left as an interesting direction for future research.

\subsection{Dual-locking Robustness}
\label{subsec:V-F-Dual-locking robustnes}

Here, we present a summarized threat landscape along with the corresponding responses of the proposed protection mechanism, expressed through the robustness of each locking level individually and the aggregate resilience defined by the logical function Level 1 OR Level 2 (see Table \ref{Table4}).

\begin{table}[H]
\centering
\caption{Dual-locking robustness of each protection level}
\label{Table4}
\begin{tabular}{
  >{\raggedright\arraybackslash}p{0.39\linewidth}
  >{\centering\arraybackslash}p{0.04\linewidth}
  >{\centering\arraybackslash}p{0.04\linewidth}
  >{\centering\arraybackslash}p{0.11\linewidth}
  >{\centering\arraybackslash}p{0.18\linewidth}
}
\hline\hline
Attacks & Level 1 & Level 2 & Model usability & Dual-Locking robustness \\
\hline
Post-lock fine-tuning                 & Yes & Yes & No  & Yes \\
Model distillation      & No  & No  & No  & No  \\
Pruning and compression & No  & Yes & No  & Yes \\
Weight reinitialization               & Yes & No  & No  & Yes \\
Overwriting or forging                & Yes & No  & Yes & Yes \\
Key-search or brute-force             & Yes & Yes & Yes & Yes \\
Collusion or ensemble attacks         & No  & Yes & Yes & Yes \\
\hline\hline
\end{tabular}

\vspace{2mm}
\noindent\parbox{\linewidth}{\footnotesize
Description of representative attacks, impact on locked model, and evaluated robustness at each protection level (Level 1 — key-based locking; Level 2 — PIN watermarking), including model usability and the aggregated resilience under logical composition (L1 OR L2).
}
\end{table}

\textit{Post-lock fine-tuning} — The PIN watermarking stage employs the QIM method \cite{Chen2001} for embedding bits into model parameters via discrete shifts \eqref{Eq.13}, applied to spaces such as the bias vector, where gradient-based corrections are inefficient. Robustness stems from the quantization threshold $\Delta$, below which parameter perturbations do not affect the embedded signal, making removal feasible only through large-scale coefficient redistribution (\ref{subsec:V-D-PIN Watermarking}). Sparse QIM further enhances resilience by applying quantization over multiple small, hidden parameter subsets \eqref{Eq.15}, ensuring that most embedded PIN bits remain intact even after extensive fine-tuning or pruning. While stage-1 fine-tuning cannot recover the accuracy of a disabled model, overall robustness still depends on tuning intensity.

\textit{Model extraction / distillation} — our scheme does not stop a black-box extraction attack that trains a surrogate model from API queries, because such attacks do not operate on the protected weight tensors. If an attacker were to gain access to the protected model, level-1 transformations would nonetheless prevent recovery of correct outputs and thus block model usability.

\textit{Pruning / sparsification / compression} — these operations remove and reshuffle weights and biases, which both reorders indices used by the permutation key (level-1) and erases PIN bits embedded by Sparse QIM (level-2). The key K is therefore vulnerable to destruction, but the attacker’s pruned model will typically be non-functional. Sparse QIM is resilient to the deletion of individual carriers because each watermark bit is spread across the entire carrier vector; extreme compression can still eliminate enough carriers to break synchronization and extraction, in which case error-correction codes (as used for 3D geometry) are applicable \cite{Vasic2013QIM-LDPC}.

\textit{Weight reinitialization / randomization} — reinitializing or randomizing weights destroys the key K and desynchronizes PIN bit positions, yielding a complete loss of locking information. Yet such an attack yields no unlocked, usable model (only a degraded one) calling into question its practical value.

\textit{Watermark overwriting / forging} — embedding a new watermark can partially or fully destroy the original PIN, depending on the attacker’s embedding algorithm. However, without the secret key the attacker cannot make the model usable. Our blind embedding means the attacker is not even aware that a watermark exists.

\textit{Key search / brute-force} — brute-force can find short PINs, but is infeasible for encrypted key lengths beyond roughly $10^4$ (see Table \ref{Table1}), which is a standard effective regime.

\textit{Collusion / ensemble attacks} — aligning and comparing multiple keyed models may hint at the key’s existence and assist reconstruction attempts, but because watermark bits are sparsely written into bias coefficients, observed coefficient differences do not directly reveal bit values.

\subsection{PIN Robustness}
\label{subsec:V-G-PIN Robustness}

To quantify the robustness of the authorship-protection stage, endurance tests were performed on the embedded PIN watermark by analyzing variations in bias coefficients used as watermark carriers. Fine-tuning was chosen as a representative non-malicious post-lock modification, applied to watermarked transformer models based on the BERT-Tiny architecture. The experiments covered three levels of complexity and fine-tuning intensity on the SST-2 dataset \cite{Socher2013Recursive}, using the $min\_sst2\_lora\_biasall$, $p2\_approx4x$, and $sst2\_lora\_0p05\_2e-5\_ep1$ LoRA configurations.
From the tested transformer model, a NumPy 1D vector was extracted by flattening all bias tensors from the Torch representation, thereby forming the host signal $x$. Sparse QIM modulation with scalar quantizers $\boldsymbol{Q}$ \eqref{Eq.13} was then applied to this signal, using a quantization step of $\Delta=0.1$ which was previously employed in the model accuracy tests in subsection \ref{subsec:V-E-Duaal-locking Transformers}) and here serves as the distortion threshold for robustness evaluation. The resulting watermarked signal $y$ \eqref{Eq.14} was subsequently subjected to fine-tuning modifications.

The robustness of the embedded PIN watermark is expressed through the bit error rate relation $\mathrm{BER}\approx\boldsymbol{Q}(\Delta/\sigma_{proj})$ where $\sigma_{proj}$ denotes the standard deviation of parameter variations projected onto the embedding subspace. The corresponding quantization step required to meet a target reliability level is given by $\Delta_\mathrm{BER}=4\sigma_{proj}\boldsymbol{Q}^{-1}(\mathrm{BER}_{target})$, while the allowable embedding distortion for Sparse QIM with block size $T$ and sparsity ratio $\rho$ is constrained by the expected mean squared error $\mathrm{MSE}\approx\rho(\Delta^2/12T)$.

For intensive fine-tuning configuration ($p_2\approx4\times$), the statistical evaluation of bias variations across all transformer layers showed that the average absolute change $|\Delta_b|$ remained within $2.2*10^{-3}$ - $5.4*10^{-3}$, while the maximum observed change $\Delta_b^{\mathrm{max}}$ did not exceed $1.9*10^{-2}$. Within the two layers containing Sparse QIM–embedded blocks (encoder.layer.0.output.dense and encoder.layer.1.attention.output.dense), $|\Delta_b|=2.36*10^{-3}$ and $|\Delta_b|=2.96*10^{-3}$, respectively, with a corresponding projected standard deviation of $\sigma_{proj}\approx3.5*10^{-3}$. The resulting BER and MSE metrics remained well below their analytical thresholds, with the estimated bit error rate below $10^{-20}$ and the mean squared embedding distortion bounded by $1*10^{-5}$, both several orders of magnitude lower than the predefined robustness limits.

\begin{figure}[H]
\centerline{\includegraphics[width=18.5pc]{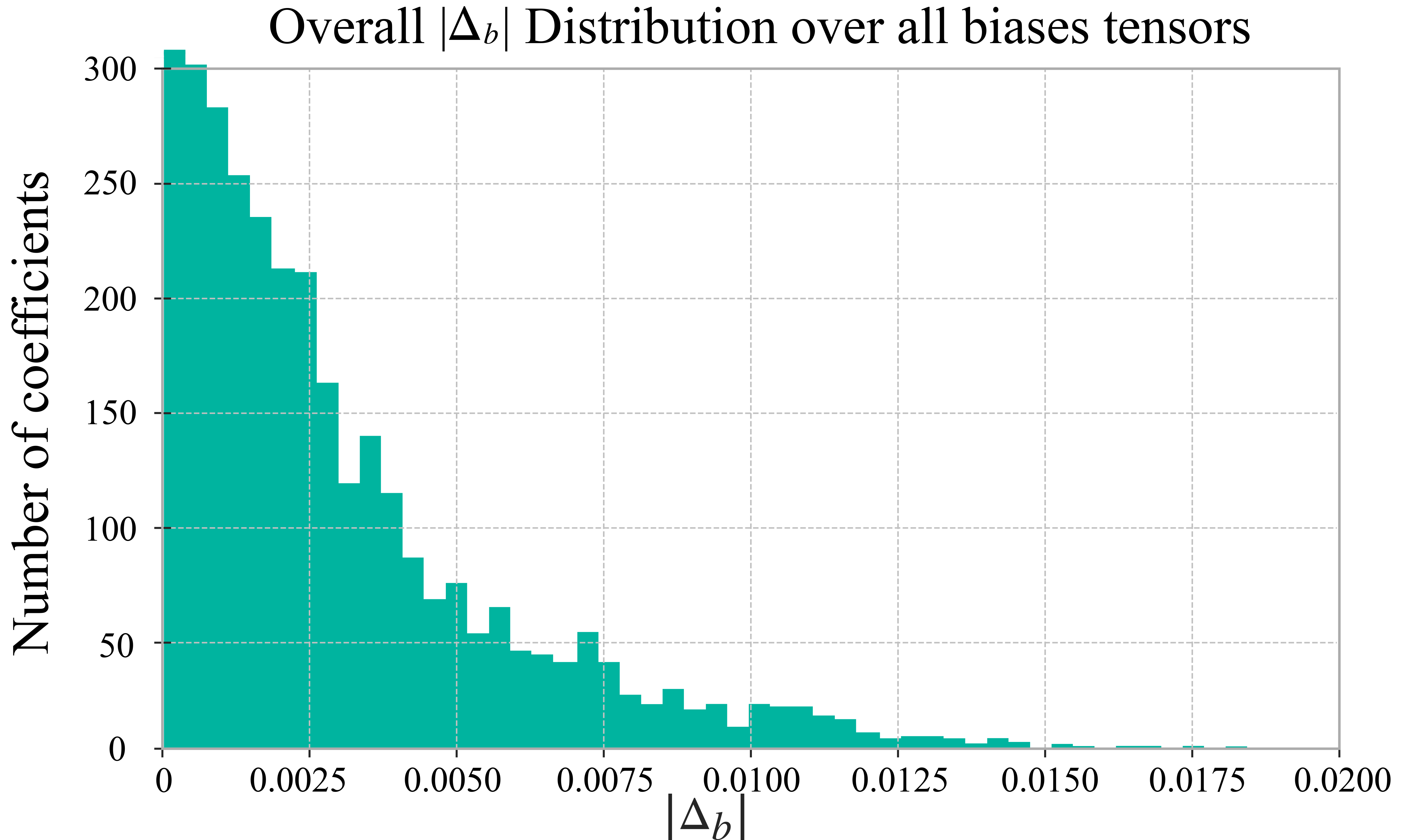}}
\caption{Histogram of absolute bias variations $|\Delta_b|$ across all Sparse QIM $T$-blocks for the intensive fine-tuning configuration ($p2\_approx4x$).}
\label{Fig.4.}
\end{figure}

The histogram (Fig. \ref{Fig.4.}.) of the $|\Delta_b|$ distribution across all $T$-blocks exhibits a pronounced concentration of values near zero, indicating minimal parameter drift during fine-tuning. Only a few sparse occurrences of larger $|\Delta_b|$ magnitudes were observed, yet all remained well below the quantization step threshold $\Delta/2=0.05$, confirming that no embedded bit approached the decision boundary and ensuring the opportunity to use an even smaller quantization step, thereby further reducing the already negligible impact of watermarking on the overall accuracy of LLM models.

\section{Conclusion}
\label{sec:VI-Conclusion}

This paper introduced a dual-locking framework for protecting neural networks, combining key-based index permutation and PIN-driven Sparse QIM watermarking. The method reshuffles indices of high-magnitude weight and bias parameters based on a cryptographic key, impairing model functionality without altering numerical values. In parallel, a blind binary watermark—derived from a user-defined Personal Identification Number (PIN)—is embedded into bias coefficients via Sparse Quantization Index Modulation. This two-stage process achieves both functional locking and cryptographic authorship verification.

Unlike perturbation-based methods, the proposed approach ensures lossless reversibility through inverse permutation, while preserving model integrity. The embedded watermark remains imperceptible during inference and provides post-hoc authentication through PIN-key association.

Implementation was fully automated in Python and evaluated on three feedforward architectures trained on MNIST, on standard CNN architectures (ResNet18, ResNet20, and ResNet32) validated on CIFAR-10/100 and ImageNet-1K, as well as on pretrained transformer models (ALBERT, BERT-Tiny, DistilBERT, and MiniLM) fine-tuned on the SST-2 dataset under three LoRA configurations. 
Experimental results demonstrate that adaptive permutations targeting high-impact weights cause sharp accuracy drops (e.g., $96.48\%$ to below $10\%$ and even $0.5\%$ for CNNs) with compact keys, clearly outperforming random reshuffling, which requires significantly longer keys ($r\geqslant10,000$). Similar behavior was observed for transformer models, further confirming the robustness and general applicability of the proposed mechanism.

Performance under different network depths, key lengths, and PIN sizes indicates that locking effectiveness depends not only on algorithmic parameters but also on training quality. In some undertrained networks, minor accuracy gains after bias reshuffling were observed, suggesting deeper interplay between training dynamics and lock resilience.

In the current version, the algorithm permits manual specification of the key length to explore the mechanism’s behavior, but also restricts it from exceeding the total number of weight coefficients. Building on these findings, the next version will enable automatic adaptation of the key length to the network architecture. Future work will also address embedding the entire key rather than only the PIN, making precise determination of key length essential to satisfying capacity requirements. Such a blind and robust watermark would provide a solution to the challenges of distribution and strong authorship verification, while the proposed mechanism remains the most effective approach for preventing unauthorized model usage.

Altogether, these findings underline the dual role of the proposed mechanism, as both a practical safeguard against unauthorized model usage and a promising foundation for developing future watermarking strategies that ensure trust, ownership, and secure distribution of neural networks.

\appendix
\label{sec:VII-Appendix}

To maintain clarity and avoid redundancy in the main body of the paper, we present here supplementary experimental results that support our analysis of the proposed neural network locking mechanism. These results provide additional empirical grounding for the claims discussed in the main text, while preserving the flow of the primary narrative by isolating dense quantitative data in a dedicated appendix.

\begin{figure}[H]
\centerline{\includegraphics[width=18.5pc]{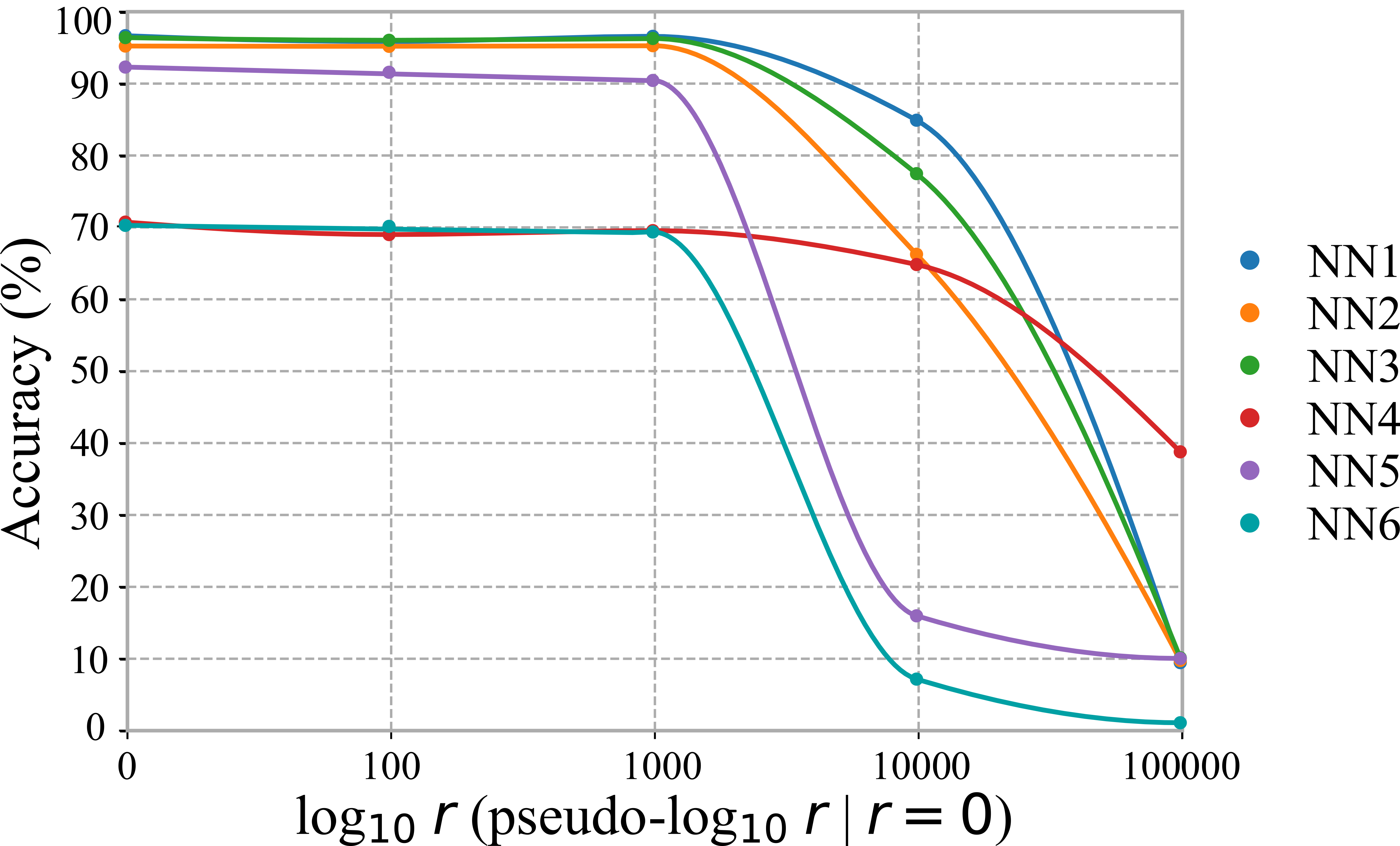}}
\caption{Accuracy degradation curves for all neural networks (NN1--NN6) as a function of key length $r$ under random reshuffling of weight indices \eqref{Eq.6}. The X-axis is represented using a base-$10$ logarithmic scale of~$r$, where the value $r = 0$ effectively corresponds to the performance of the original networks.}
\label{Fig.5.}
\end{figure}

The previous figure (Fig. \ref{Fig.5.}.) demonstrates that, in the case of random reshuffling, the degradation effects on accuracy become evident only for large key lengths.

This indicates that effective locking, manifested as an accuracy reduction to approximately $10\%$ , can be achieved only through random reshuffling of nearly all weight coefficients in the network. The accuracy test of NN3, the deepest architecture with the highest neuron density per layer, further supports this conclusion: its accuracy did not fall below $63.04\%$ even for a key length of $r = 10^5$. These results align with the expected behavior of the proposed locking scheme, which was designed to exhibit significant degradation only under large-scale random reshuffling. Interestingly, an anomalous accuracy increase was observed in larger networks for shorter key lengths, likely reflecting insufficient baseline optimization of the unlocked model ($93.29\%$).

Table \ref{Table5} shows the outcomes of applying random index permutation to the bias vectors of the neural networks (Fig. \ref{Fig.2.}.). Although bias coefficients are critical to neuron activation, the effect of reshuffling these values alone was surprisingly minimal. 

\setcounter{table}{0}
\renewcommand{\thetable}{\thesection.\Roman{table}}
\begin{table}[H]
\centering
\caption{Accuracy of Neural Networks}
\label{Table5}
\begin{tabular}{>{\raggedright\arraybackslash}p{0.2\linewidth}>{\centering\arraybackslash}p{0.12\linewidth}>{\centering\arraybackslash}p{0.12\linewidth}>{\centering\arraybackslash}p{0.12\linewidth}>{\centering\arraybackslash}p{0.12\linewidth}}
\hline\hline
 {NN Name} & \multicolumn{4}{c}{Locked NN Accuracy for different key lengths} \\ 
 & {Original} & {10} & {100} & {$>100$} \\ 
 \hline
B\_NNetwork1    & 96.48\% & 96.26\% & 95.98\% & 95.47\% \\ 
B\_NNetwork2    & 95.04\% & 95.02\% & 95.06\% & 93.44\% \\ 
B\_NNetwork3    & 96.21\% & 96.24\% & 95.84\% & 94.86\% \\ 
ResNet18 & 70.59\% & 52.96\% & 0.42\% & 0.38\% \\
ResNet20 & 92.12\% & 75.31\% & 10.04\% & 10.00\% \\
ResNet32 & 70.14\% & 55.70\% & 7.09\% & 1.16\% \\
\hline\hline
\vspace{3mm}
\end{tabular}
\noindent\parbox{\linewidth}{\footnotesize The outcomes of applying random index permutation to the bias vectors of the neural networks.}
\end{table}

Table \ref{Table6} provides results from the adaptive bias reshuffling procedure, analogous to the adaptive weight-locking method. Here, the reshuffling targets biases with the highest and lowest absolute values, forming index-pair keys designed to disrupt neuron-level behavior selectively. Despite this targeted approach, the degradation effect remained weaker than that of weight-based locking, particularly for networks with original accuracy below 95\%. In fully connected architectures, the effect was more visible, while for CNNs such as ResNet18 it dropped to only $0.16\%$ even for key lengths of $r$ = 10.

\begin{table}[H]
\centering
\caption{Accuracy of Neural Networks: adaptive locked biases}
\label{Table6}
\begin{tabular}{>{\raggedright\arraybackslash}p{0.2\linewidth}>{\centering\arraybackslash}p{0.12\linewidth}>{\centering\arraybackslash}p{0.12\linewidth}>{\centering\arraybackslash}p{0.12\linewidth}>{\centering\arraybackslash}p{0.12\linewidth}}
\hline\hline
{NN Name} & \multicolumn{4}{c}{Locked NN Accuracy for different key lengths} \\ 
& {Original} & {10} & {100} & {$>100$} \\ 
\hline
B\_NNetwork1 & 96.48\% & 94.38\% & 95.10\% & 96.48\% \\ 
B\_NNetwork2 & 95.04\% & 94.49\% & 90.82\% & 95.04\% \\ 
B\_NNetwork3 & 96.21\% & 95.68\% & 93.58\% & 96.21\% \\
ResNet18 & 70.59\% & 0.16\% & 0.06\% & 0.22\% \\
ResNet20 & 92.12\% & 9.63\% & 10.15\% & 10.00\% \\
ResNet32 & 70.14\% & 1.03\% & 1.07\% & 1.00\% \\
\hline\hline
\vspace{3mm}
\end{tabular}
\noindent\parbox{\linewidth}{\footnotesize The outcomes of applying adaptive index permutation to the bias vectors of the neural networks.}
\end{table}

The results reveal a limited and highly network-dependent impact on model accuracy. An intriguing pattern emerges in Table \ref{Table6}, where certain networks such as NN2 initially experience accuracy degradation with increasing key length (e.g., from $r=10$ to $r=100$), followed by partial recovery once the key length exceeds a specific threshold. This non-monotonic trend resembles the behavior previously observed in deeper networks under random reshuffling, where moderate perturbations disrupt layer dynamics, while extensive reshuffling induces new stable patterns that the model unexpectedly aligns with latent representations.

Although the adaptive bias reshuffling (see Table \ref{Table6}) is more selective than random reshuffling, the observed accuracy rebound indicates complex interactions between bias perturbations and internal compensation mechanisms, particularly in undertrained or sub-optimally structured networks. This observation supports the earlier hypothesis that fragile architectures exhibit unstable responses to partial perturbations, whereas full reshuffling either overwhelms or re-stabilizes specific layers.

For convolutional architectures, the effect becomes both clearer and more extreme. As shown in Tables \ref{Table6}–\ref{Table7}), ResNet18 collapses to only 0.16\% accuracy at $r=10$ under adaptive reshuffling, while random reshuffling already reduces accuracy from 70.59\% to 52.96\% for the same key length and below 1\% for longer keys. ResNet20 and ResNet32 display a similar saturation effect, converging to near-random accuracy ($\approx10\%$) once $r>10$. This confirms that CNN bias vectors, defined per feature map rather than per neuron, constitute a compact parameter set whose reshuffling disrupts global activation balance and normalization across filters, rapidly degrading representational stability.

This asymmetric sensitivity, moderate in fully connected networks and catastrophic in CNNs, suggests that PIN embedding within bias structures must strictly avoid long PIN lengths, especially considering that binary watermark encoding inherently demands higher channel capacity. In practice, PIN lengths above $10$ should be excluded, and even shorter keys are preferable, despite the fact that the actual PIN embedding via QIM quantization introduces coefficient value modifications far milder than any reshuffling process. Consequently, bias parameters remain suitable for lightweight, low-intensity PIN watermarking, but not for reshuffling-based locking or dense key-embedding schemes.

\begin{table}[H]
\centering
\caption{Accuracy of watermarked LLMs}
\label{Table7}
\begin{tabular}{>{\raggedright\arraybackslash}p{0.2\linewidth}>{\centering\arraybackslash}p{0.10\linewidth}>{\centering\arraybackslash}p{0.10\linewidth}>{\centering\arraybackslash}p{0.10\linewidth}>{\centering\arraybackslash}p{0.10\linewidth}}
\hline\hline
{LLMs} & \multicolumn{4}{c}{Watermarked Accuracy for different key lengths}\\
 & Original & 1000 & 10000 & 15000 \\
\hline\hline
ALBERT & 92.66 & 82.68 & 50.46 & 50.69 \\ 
BERT-Tiny & 70.59 & 79.93 & 59.29 & 49.43 \\ 
MiniLM & 91.06 & 89.45 & 49.77 & 48.97 \\ 
MiniLM & 90.14 & 87.96 & 57.80 & 57.80 \\  
\hline\hline
\vspace{3mm}
\end{tabular}
\noindent\parbox{\linewidth}{\footnotesize The binary decision for a specific transformers architecture after applying the adaptive locking mechanism with different key lengths $r$. Dual-locking mechanism is performed in whole list of embeddings.}
\end{table}

The Table \ref{Table7} presents the accuracy of transformer models when locking the entire set of embedding tensors, in contrast to the dual-locking effects applied exclusively to the word embedding layer as summarized in \ref{Table3}.

Comparison of Table \ref{Table3} and Table \ref{Table7} shows that locking all embedding tensors (word, positional, and segment) causes slightly stronger degradation and faster convergence to the 50\% baseline, confirming full disruption of semantic coherence. After normalization to the random-guess level, the residual effective accuracy remains below $10\%$ for all models, indicating complete functional locking and consistent robustness of the proposed mechanism.

\section*{Acknowledgment}
This work was supported by the \textit{Smart Living Lab} (\url{https://www.smartlivinglab.ch/en/}), a joint project funded by the University of Fribourg, EPFL, and HEIA-FR.

The authors would like to thank OpenAI’s ChatGPT \cite{Openai2025chatgpt} for assistance with language editing, translation, and improving text clarity during manuscript preparation.

\bibliographystyle{IEEEtran}
\bibliography{References}

\begin{IEEEbiography}[{\includegraphics[width=1in,height=1.25in,clip,keepaspectratio]{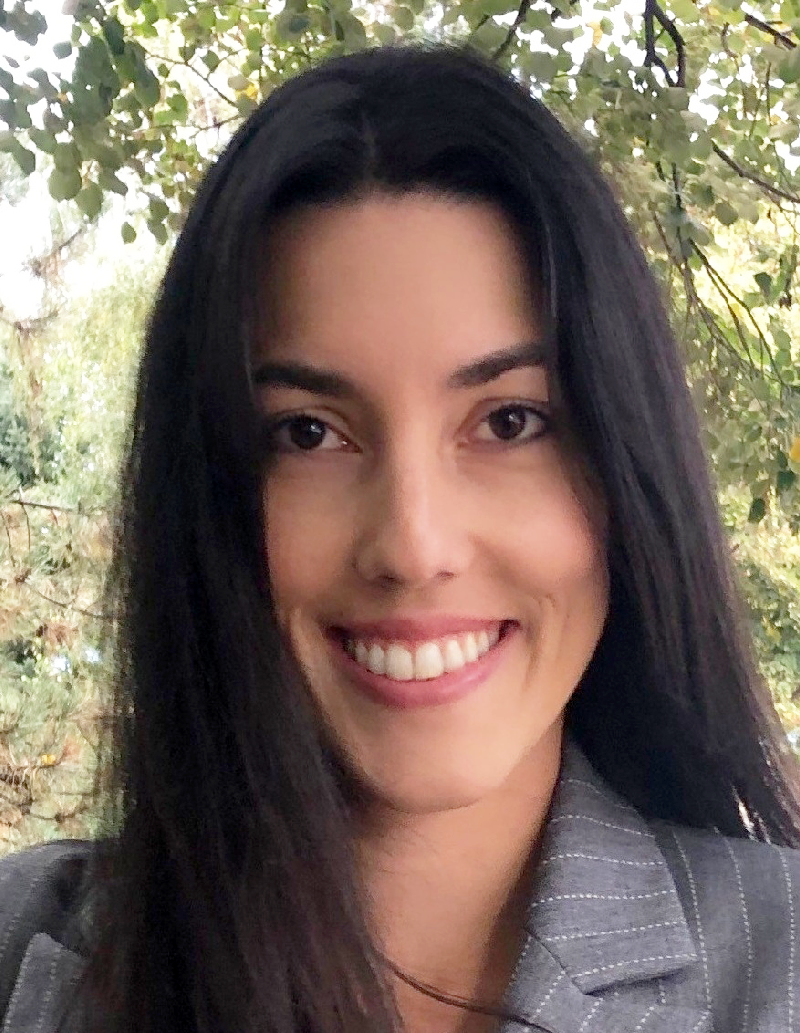}}]{Iva Vasic}(S’24) was born in Nis, Yugoslavia (now Serbia). In 2025, she received a PhD degree in the field of Engineering from the Polytechnic University of Marche, Ancona, Italy. She is currently a postdoctoral researcher in Computer Science at the Faculty of Science and Medicine, University of Fribourg, Fribourg, Switzerland.

She teaches the master's level course "Foundations of Spatial Computing and Applications in Augmented and Virtual Reality" at the Faculty of Management, Economics and Social Sciences, University of Fribourg, and serves as a coach for the project "Collaborative Stationary Robot for Smart Living Consumer Use Cases using Spatial Conceptual Modeling," supported by the Smart Living Lab Student Innovation Grants. Since her enrollment in doctoral studies in 2021, she has been a member of the Vision, Robotics and Artificial Intelligence (VRAI) research team (Italy). In 2023, she was also a visiting PhD student at the Digitalization and Information Systems (DIGITS) Group at the University of Fribourg, Switzerland, where she conducted research focused on artificial intelligence (AI), knowledge graphs, and three-dimensional (3D) visualization.

Dr. Vasic was involved in the Virtual Immersion in Territorial Arts (V.I.T.A.) project from 2021 to 2024, in the Digital Curator Training \& Tool Box (DCbox) Erasmus+ Programme project from 2022 to 2024, and in the INTERREG - IPA in 2013 and 2020. She has served as a reviewer for multiple venues, including the ACM and Expo DH25. She won the Salento AVR 2021 Best Paper award. Her research interests include large language models, spatial computing, neural and neuro-symbolic AI, human-computer interaction, and mathematical approaches to 3D geometry understanding.
\end{IEEEbiography}

\begin{IEEEbiography}[{\includegraphics[width=1in,height=1.25in,clip,keepaspectratio]{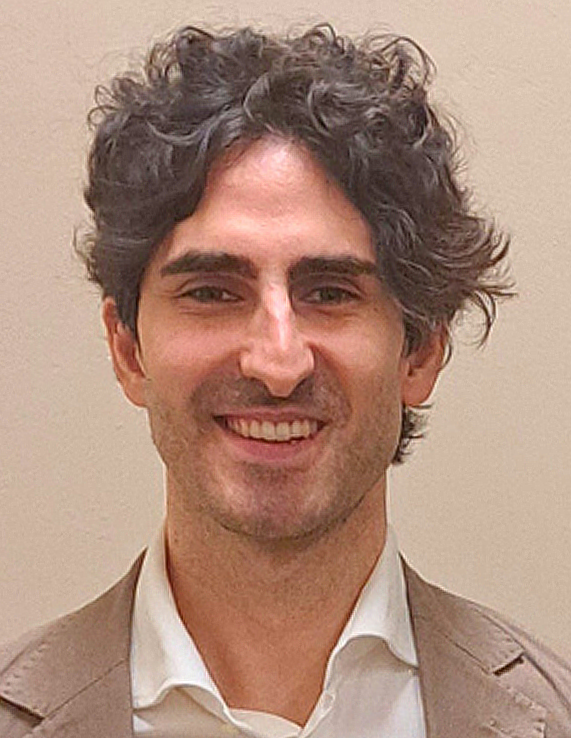}}]{Jesús Muñoz-Cádiz} was born in Osuna, Sevilla, Spain. He received a PhD at the Università Politecnica delle Marche, Ancona, Italy. He is a postdoctoral researcher at the University of Fribourg, in the Faculty of Science and Medicine (Fribourg, Switzerland). In late 2023, he was a visiting PhD student at the University of the Basque Country, Spain, where he gained expertise in data structures within the Building Information Modeling (BIM) methodology, including the architecture of the Industry Foundation Classes (IFC) metadata schema.

His research formulates the three-dimensional (3D) reconstruction of complex geometries as an inverse problem on discrete manifolds, integrating differential geometry operators within Poisson Surface Reconstruction to enforce topological consistency. He has devised unsupervised segmentation pipelines for point clouds, combining geometric feature descriptors with K-Means clustering and robust plane detection. His experience with ontological frameworks for BIM enabled him to develop the A$^{2}$Heritage, a native IFC library for structured semantic annotation in the built heritage domain.
    
Dr. Muñoz-Cádiz current research explores Large Language Models (LLMs) into software engineering workflows, with special emphasis on AI agent architectures for code completion and generation within open-source ecosystems. He teaches the master's level course "Foundations of Spatial Computing and Applications in Augmented and Virtual Reality" at the Faculty of Management, Economics and Social Sciences, University of Fribourg. Additionally, he is the coach for the project "Implementation of Construction Process Modeling Language in MMAR" supported by the Smart Living Lab Student Innovation Grants.
\end{IEEEbiography}

\newpage
\begin{IEEEbiography}[{\includegraphics[width=1in,height=1.25in,clip,keepaspectratio]{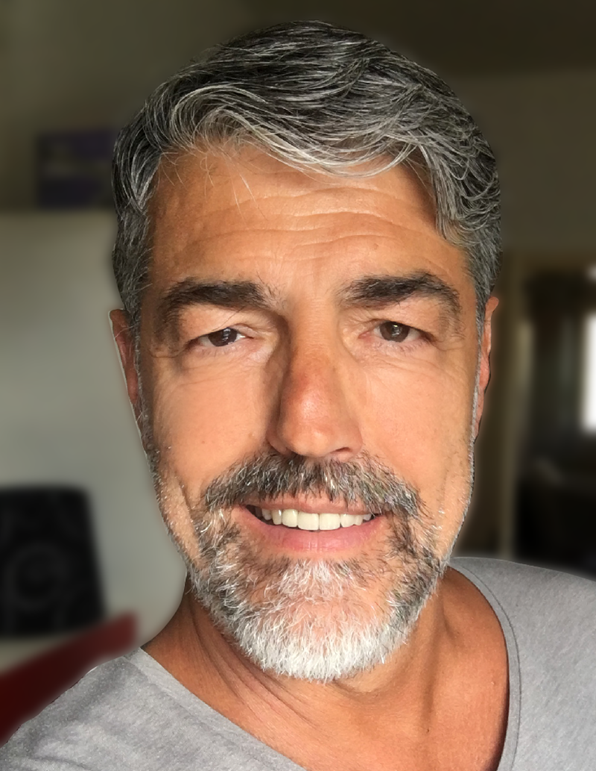}}]{Bata Vasic} was born in Bela Palanka, Yugoslavia (now Serbia). He received the B.Sc. and M.Sc. degrees in electrical engineering and telecommunications from the University of Niš, Serbia, in 1992 and 1993, respectively, and the Ph.D. degree in electrical engineering and computer science from the University of Niš, Faculty of Electronic Engineering in 2013.

In 1992, he pioneered architectural 3D visualization in the Balkans through early research on special effects for film. Since 2008, he has taught courses in computer animation, computer graphics, and computer design at the Department of Electronics, Faculty of Electronic Engineering, University of Niš. Since 2017, he has also served as an Assistant Professor in the Department of Information Technology at MB University, PPF Belgrade. In 2022, as an Associate Professor, he began teaching courses in multimedia, web application development, computer graphics, and information theory and secure coding at undergraduate and master’s levels, as well as artificial intelligence, cryptography, and watermarking at the doctoral level. Concurrently, as a Research Professor at the University of Niš, he leads two EU-funded research projects focused on AI development and applications.
    
Prof. Vasić was invited in 2015 as a Visiting Professor at ENSEA, the Faculty of Electrical Engineering at the University of Cergy-Pontoise, France, working within the ETIS Laboratory for Image Processing. In 2013, he served as an external evaluator for the diploma committees of the Moving Images program at MCAST—the Faculty of Arts, Science, and Technology in Malta. He has been engaged on multiple occasions as a reviewer for several leading scientific journals, including IEEE Transactions on Medical Imaging, IEEE Transactions on Multimedia, IEEE Transactions on Visualization and Computer Graphics, Elsevier Signal Processing, ACM Journal on Computing and Cultural Heritage, and Multimedia Tools and Applications. His research interests include neural network architectures, features of large language models, multimodal AI development, and their application and protection, as well as 3D mesh geometry with a focus on geometric feature-based watermarking. He is also involved in medical imaging research, including automatic segmentation and the automatic construction of 3D medical geometries.
\end{IEEEbiography}

\end{document}